\documentclass[aps,twocolumn,floatfix,superscriptaddress]{revtex4}

\usepackage{amssymb,amsmath,amstext}                
\usepackage{graphicx}                                               
\usepackage{epstopdf}                                               
\usepackage{color}                                                     
\usepackage{bm}                                                        
\usepackage{appendix}                                              
\usepackage{bbold}
\usepackage{bbm}
\usepackage{ulem}
\usepackage{latexsym}
\usepackage[colorlinks=true,citecolor=blue,linkcolor=magenta]{hyperref}
\usepackage{cancel}
\usepackage{relsize}

\usepackage{soul}
\usepackage{mathtools}

\def\be{\begin{equation}}
\def\ee{\end{equation}}
\def\bea{\begin{eqnarray}}
\def\eea{\end{eqnarray}}
\def\ra{\rangle}
\def\la{\langle}
\def\bi{\begin{itemize}}
\def\ei{\end{itemize}}

\definecolor{dgreen} {RGB}{78,138,21}

\begin{document} 

\title{Nonlinear entanglement growth in inhomogeneous spacetimes}

\author{Arkadiusz Kosior} 
\affiliation{Max-Planck-Institut f\"ur Physik Komplexer Systeme,
N\"othnitzer Strasse 38, D-01187, Dresden, Germany}
\author{Markus Heyl} 
\affiliation{Max-Planck-Institut f\"ur Physik Komplexer Systeme,
N\"othnitzer Strasse 38, D-01187, Dresden, Germany}

\date{\today}

\begin{abstract}
Entanglement has become central for the characterization of quantum matter both in and out of equilibrium. In a dynamical context entanglement exhibits universal linear temporal growth in generic systems, which stems from the underlying linear light cones as they occur in planar geometries. Inhomogeneous spacetimes can lead, however, to strongly bent trajectories. While such bent trajectories crucially impact correlation spreading and therefore the light-cone structure, it has remained elusive how this influences the entanglement dynamics. In this work we investigate the real-time evolution of the entanglement entropy in one-dimensional quantum systems  after quenches which change the underlying spacetime background of the Hamiltonian. Concretely, we focus on the Rindler space describing the spacetime in close vicinity to a black hole. As a main result we find that entanglement grows sublinearly in a generic fashion both for interacting and noninteracting quantum matter. We further observe that the asymptotic relaxation becomes exponential, as opposed to algebraic for planar Minkowski spacetimes, and that in the vicinity of the black hole the relaxation time for large subsystems becomes independent of the subsystem size. We study entanglement dynamics both for the case of noninteracting fermions, allowing for exact numerical solutions, and for random unitary circuits representing a paradigmatic class of ergodic systems.

\end{abstract}
\date{\today}

\maketitle

\section{Introduction}
Quantum entanglement \cite{bengtsson_zyczkowski_2006} is a key feature of quantum systems and has become a cornerstone of many modern branches of physics, such as quantum information \cite{Horodecki2009} (in particular, quantum cryptography \cite{Ekert1991,Gisin2002,Pirandola2019}, quantum  teleportation \cite{Bennett1993,Pirandola2015}, quantum  computing \cite{Feynman1982,Steane1998,Ladd2010,Hauke2012,Georgescu2014}) and quantum many-body physics \cite{Amico2008,Laflorencie2016}.  In the latter, the importance of entanglement is reflected by its universal properties, both under equilibrium and non-equilibrium conditions. Ground states of gapped one-dimensional quantum systems obey an area law  \cite{Holzhey1994,calabrese2004entanglement,Eisert2010}. In non-equilibrium the entanglement entropy of generic systems shows generally a linear growth in time, which is today explained through a semiclassical picture of ballistic quasiparticle propagation \cite{Calabrese2005,Fagotti2008,Nezhadhaghighi2014,Bucciantini_2014,Cardy_2016,cotler2016entanglement,Wen_2018,Najafi2018,Kim2013,Buyskikh2016,Alba2017,Alba2018,Alba2018b,Bertini_2018}. Although the applicability of the quasiparticle picture has been confirmed for a range of integrable models, including inhomogeneous initial states \cite{Viti_2016,Dubail2017,Bertini2018b,Alba2019}, this picture cannot be directly applied when the post-quench Hamiltonian is not translationally invariant. A prominent example is the logarithmic spreading of entanglement in a many-body localized phase   \cite{Bardarson2012,serbyn2013universal,huse2014phenomenology,Schreiber2015,Abanin2019} in disordered models. These models are, however, constructed on flat homogeneous spacetime backgrounds and therefore their local properties are the same everywhere. It has remained, however, an open question how entanglement grows in spatially inhomogeneous systems due to inhomogeneity of the spacetime itself.

In this work we investigate the entanglement entropy growth generated by a quench of the underlying spacetime metric of the Hamiltonian. As a prototypical example of an inhomogeneous spacetime we choose a Rindler space, which can be viewed an asymptotic spacetime near a black hole horizon \cite{birrell_davies_1982,takagi1986vacuum,wald2010general}. We find that in the case of a global inhomogeneous quench the initial entanglement entropy growth is sublinear and attains a constant size-independent value in a long-time limit. This is the most noticeable difference in comparison to the translationally invariant case where the entanglement entropy growth is linear.  We contend that this behavior is  universal applying for both integrable and ergodic systems.  We find that some important aspects of the physical picture can be captured qualitatively via semiclassical arguments based on the maximal speed of correlation propagation. The correlations spread within bent light-cones defined by null geodesic of the (1+1) Rindler metric which suggests that our results capture the universal properties of the continuum theory. 

This papers is organized as follows. In Sec.~\ref{sec:model} we discuss the basic concepts which are central to this study and introduce the model. In Sec.~\ref{sec:correlations} we calculate the dynamics of correlations after a Rindler quench and draw semiclassical arguments that are helpful to understand some physical aspects of the entanglement evolution. In Sec.~\ref{sec:numerics} we present numerical results for finite subsystems and, by achieving a data collapse, we identify size-independent universal behavior of entanglement. While in Sec~\ref{sec:model}~-~\ref{sec:numerics} we focus on the case of free fermions on the 1D lattice, Sec.~\ref{sec:randomcircuits} is devoted to random unitary circuits \cite{nahum2017quantum,jonay2018coarsegrained,Nahum2018,Nahum2018b} that represent minimally structured ergodic models. We estimate the coarse grained entanglement entropy, which is equivalent to the minimal membrane description \cite{jonay2018coarsegrained}, and show the remarkable qualitative agreement of results for integrable and ergodic systems. In Sec.~\ref{sec:conclusions} we conclude. 

\section{The model and setup}\label{sec:model}
\subsection{Entanglement entropy}\label{sec:motivation}

Quantum entanglement has developed into a central concept in quantum many-body physics.
One of the key consequence for a state $|\Psi\ra$ to be entangled is that it cannot be written as a simple product of states belonging to different subsystems. Quantification of entanglement is possible through various entanglement measures \cite{plenio2014introduction}.
In this work we want to study the entanglement entropy defined as 
\be\label{ee}
S= -\mbox{Tr}_A\left[\hat \rho_A \ln \hat \rho_A \right],
\ee
where $\hat \rho_A= \mbox{Tr}_B |\Psi\ra\la\Psi|$  is a reduced density matrix of a subsystem $A$ traced over the rest of the system, a subsystem~$B$.

Entanglement between two spatial intervals can be generated dynamically through a quantum quench when a system is prepared in an initial state which is not an eigenstate of the post-quench Hamiltonian. Calabrese and Cardy \cite{Calabrese2005} have shown by path integral methods of (1+1) quantum field theory that, up to a time $t^* \propto L/2$, the entanglement entropy growths linearly with a rate independent of the subsystem size $L$ and  attains a constant value $\propto L$  afterwards, yielding a volume law, which is due to the finite subsystem size.

The linear spreading of entanglement entropy turns out to be very general and holds in various non-interacting \cite{Fagotti2008,Nezhadhaghighi2014,Bucciantini_2014,Cardy_2016,cotler2016entanglement,Wen_2018,Najafi2018} and short-range interacting \cite{Kim2013,Buyskikh2016,Alba2017,Alba2018,Alba2018b} models, also for inhomogeneous initial states \cite{Viti_2016,Dubail2017}. (However, deviations from the linear growth can be found in a presence of Markovian bath \cite{Maity2020}.) The linear growth of entanglement is due to the maximal speed of information. In the (1+1) Minkowski spacetime 
\be\label{minkowski_metric}
\mathrm{d}s^2= -c^2 \,\mathrm{d}t^2 + \mathrm{d}x ^2,
\ee
the rate of any information propagation is bounded by the speed of light~$c$ and all particle trajectories lie within a light-cone defined by null geodesic. In the lattice systems, the role of the speed of light $c$ plays the Lieb-Robinson velocity \cite{lieb1972finite}, being the emergent maximal velocity of correlation spreading.

\subsection{Rindler metric}

Although the entanglement dynamics in spatially invariant quantum systems seems to be completely understood, the entanglement dynamics in inhomogeneous spacetimes is a completely open question. (However, driven inhomogeneous systems have been already investigated \cite{wen2018,Lapierre2020}.)
In this paper, we study the entanglement evolution in systems on a Rindler background described by a (1+1) metric \cite{birrell_davies_1982,takagi1986vacuum,wald2010general}
\be\label{rindler_metric}
\mathrm{d}s^2= -x^2 \,\mathrm{d}t^2 + \mathrm{d}x ^2, 
\ee
which serves us as a nontrivial example of an inhomogeneous metric.  In the literature the Rindler metric appears primarily in two related contexts. First of all, the Rindler metric is an asymptotic Schwarzschild metric \cite{schwarzschild1916gravitational} in the vicinity of a black hole horizon at $x=0$. Secondly, the Rindler metric describes a flat Minkowski spacetime in a hyperbolically accelerated reference frame, i.e., it characterizes the motion of uniformly accelerated observers \cite{Unruh1976}.

Comparing Minkowski and Rindler metrics, one could interpret $c(x)=x$ as a spatially varying speed of light. Indeed, the geodesics equations come down to a simple $\mathrm{d}x/\mathrm{d}t=\pm x$ with a straightforward solution
\be\label{geodesics}
x(t)= x_0 e^{\pm t},
\ee
with $x_0$ the initial position. When one switches to a proper time $\tau=x_0 t/c$ of an observer at rest at $x=x_0$, then 
\be\label{eq:geodesics_rindler}
x(\tau)= x_0 \exp(\pm c \tau/x_0).
\ee
The above result means that the light-cones in the Rindler spacetime are distorted and the strongest bending takes place in a vicinity of a horizon at $x=0$. Conversely, the bending of the light cone is negligible for $\tau/x_0 \ll 1$, i.e., at very early times or far away from the horizon.

Because the information in the Rindler spacetime does not propagate linearly, it is a central open question how the entanglement is propagating in such a setup. In the following we argue the entanglement entropy in generic quantum systems in a Rindler spacetime first grows sublinearly and asymptotically at long-times attains a constant size-independent value in a thermodynamic limit.

\subsection{Setup}

In the following we will consider the Hamiltonian description of free fermions in inhomogeneous spacetimes. As a first step we investigate (1+1) dimensional metric, but the generalization to higher dimensions is possible, and we plan to purse this path in the next future. 
Let us consider a system of spinless fermions on a one dimensional (1D) lattice of length $N$ with open boundary conditions. We assume that an initial state is a spatially homogeneous product state
\be\label{initial_state}
|\Psi_0\ra = \Pi_{n=1}^{N/2} \hat c_{2n}^\dagger|0\ra,
\ee
where $|0\ra$ is a particle vacuum state, i.e., it is annihilated by any annihilation operator $\hat c_n |0\ra =0 $. Since the initial state is a product state, its entanglement entropy, Eq.~\eqref{ee}, vanishes with respect to any bipartition. Suppose that at time $t=0$ we perform a quench such that the evolution of a system $t>0$ is described by a new Hamiltonian
\be\label{postquench_hamiltonian}
\hat H=-\frac{1}{2} \sum_{n=1}^{N} t_n \hat c_{n+1}^\dagger \hat c_{n} + H.c.,
\ee
where $c_{n}^\dagger$, $c_{n}$ are standard fermionic creation and annihilation operators on a lattice.
It has been shown in Ref.~\cite{Boada_2011} (see also Refs~\cite{celi2017different,rodriguez2017synthetic,kosior2018,louko2018thermality,Yang2020}) that the massless Dirac fermions propagation on static spacetimes can be described by a lattice free fermion Hamiltonian with a spatially varying tunneling amplitude proportional to the determinant of the metric. For the non interacting case the spin degrees of freedom separate and therefore
 the Hamiltonian~Eq.~\eqref{postquench_hamiltonian}  corresponds to a spinless version of the discrete relativistic Hamiltonian on a curved (1+1) dimensional spacetime. 
Here, we choose $t_n=c=1$  or $t_n = n$. The first choice of tunneling amplitudes amounts to the Minkowski metric, Eq.~\eqref{minkowski_metric}, and the corresponding Hamiltonian is obviously a free fermionic Hamiltonian. From now on, we will refer to this choice of tunneling amplitudes as the homogeneous quench. The second choice of tunneling amplitudes ($t_n = n$) entails the Rindler metric, Eq.~\eqref{rindler_metric}, yielding the Rindler Hamiltonian 
\be\label{rindler_hamiltonian}
\hat H=-\frac{1}{2} \sum_{n=1}^{N} n \,\hat c_{n+1}^\dagger \hat c_{n} + H.c. \,.
\ee
We note that this Hamiltonian, in a context of the Bisognano-Wichmann theorem, can be also interpreted as a modular or entanglement Hamiltonian. The study of entanglement Hamiltonian for lattice models has recently attracted substantial attention   \cite{dalmonte2018quantum,Giudici2018,Eisler_2018,Parisen2018,Zhu2019,Turkeshi2019,MendesSantos2019,Mendes_Santos_2020,zhang2020lattice}. 

In this work, we investigate the entanglement evolution generated by the Rindler quench. In Sec.~\ref{sec:numerics} we present the exact numerical diagonalization results for the entanglement entropy and in Sec.~\ref{sec:randomcircuits} we consider an analogous quench protocol in a unitary circuit setup. Before going to the main results, in Sec.~\ref{sec:geometric_picture} for illustrative and comparison purposes we review the semi-classical pair quasiparticle picture for the homogeneous quench, which is well understood in the existing literature \cite{Calabrese2005,Fagotti2008,Nezhadhaghighi2014,Bucciantini_2014,Cardy_2016,cotler2016entanglement,Wen_2018,Najafi2018,Kim2013,Buyskikh2016,Alba2017,Alba2018,Alba2018b,Bertini_2018,Viti_2016,Dubail2017}, and in Sec.~\ref{sec:correlations} we give its geometric interpretation in terms of propagation of correlations. On the contrary, because of the lack of the translational invariance in the Rindler Hamiltonian, Eq.~\eqref{rindler_hamiltonian}, the (quasi-)momenta are no longer good quantum numbers and the quasiparticle picture of counterpropagating pairs with opposite momenta does not hold. Nevertheless, since the speed of correlation spreading is still bounded in Sec.~\ref{sec:correlations} we present heuristic arguments that allow us to extract important features of the entanglement dynamics, such as sublinear growth and a long-time asymptotically constant behavior.  Most of all, we show that this approach allows us to identify relevant scaling parameters of the model.

\subsection{Entanglement entropy for free fermionic systems: the Peschel formula}\label{sec:peschel_method}

For the free fermionic models the entanglement entropy, between a subsystem $A$ and its complement $B$, can be efficiently calculated via correlation functions \cite{Peschel_2003,Peschel_2009}, as long as an initial state can be described by a Slater determinant. In this case, 
the density matrix can be written as an exponential of free fermionic operators
\be\label{reduced_density_mtrx}
\hat \rho_A= \frac{1}{Z} e^{-\mathcal{\hat{H}}_A}, \quad \hat{H}_A= \sum_{i,j\in A}h_{ij} \hat c_i^\dagger \hat c_j, 
\ee
where $H=[h_{ij}]_{i,j\in A}$ matrix is diagonalized by the same transformation as a single particle correlation matrix $C=[\la \hat c_i^\dagger \hat c_j \ra]_{i,j\in A}$.  It has been found that the two matrices are related~\cite{Peschel_2003}
\be \label{h-c-relation}
H=\ln [(1-C)/C].
\ee
Using the thermal form of a reduced density matrix, Eq.~\eqref{reduced_density_mtrx}, one can readily obtain an expression for the entanglement entropy following a quench at $t=0$
\be\label{ee_peschel_formula}
S(t)= - \sum_{n=1}^{L} \bigg[ \lambda_n(t)\ln\lambda_n(t) +\big[1-\lambda_n(t)\big] \ln \big[1-\lambda_n(t)\big]   \bigg],
\ee
where $\lambda_n(t) $'s are eigenvalues of equal time correlation matrix   

\be
C (t) = \left[ \la \hat c_j^\dagger (t) \hat c_k (t)\ra \right]_{j,k\in A}
\ee
restricted to a subsystem $A$.  According to Eq.~\eqref{ee_peschel_formula} the maximal value of the entanglement entropy  $S_{\mathrm{max}}=L \ln 2 $ is obtained if all $\lambda_n = 1/2$, which corresponds to a trivial reduced density matrix, Eq.~\eqref{reduced_density_mtrx}, i.e., an infinite temperature state of a subsystem. 

\begin{figure}[bt]
\includegraphics[width=1\columnwidth]{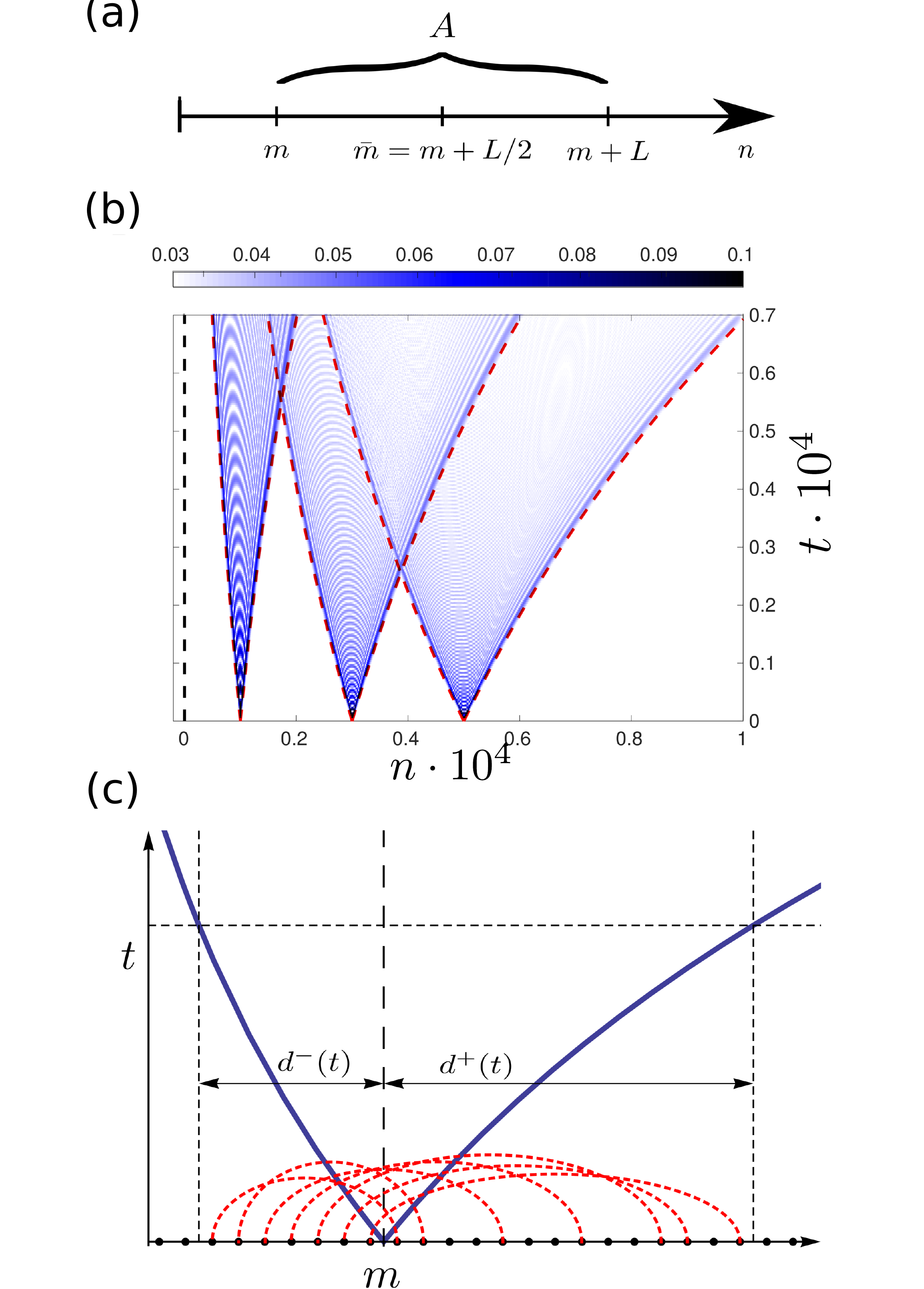}  
\caption{Panel (a): We assume a bipartition where a smaller subsystem $A=[m,m+L)$ is a segment with its mean position $\bar m = m +L/2$.
Panel (b): The exact numerical lattice calculations  of the correlation function 
  $C_{n,n'}(t) = \la \hat c^\dagger_n(t) \hat c_{n'} \ra $, see Eq.~\eqref{correlation_funtion}, followed by a Rindler quench at $t=0$. We plot $C_{n,n'}(t)$ as a function of $n$ and $t$ for different values of $n'$. We observe a non-linear spreading of correlations which form distorted light-cone structure. The light cone edges (red dashed lines) are described by a continuum theory, see Eq.~\eqref{geodesics}. 
  Panel (c): Because of the light-cone bending, the number of lattice sites that are in a causal relation with a site $m$ depend on a spatial direction. We assume that the entanglement entropy is equal to the number of distinct pairs that can be correlated across the boundary between $A$ and $B$, i.e., it is proportional to $\min(d^-,d^+) = d^-$. 
}
\label{fig:light-cone}   
\end{figure}

Throughout this article we consider a bipartition $A$ and $B$, where $A=[m,m+L)$ is the smaller subsystem of length $L$ and $B$ is its complement, see Fig.~\ref{fig:light-cone}~A. Also, by $\bar m = m+L/2 $ we denote the  position of the middle of  subsystem $A$, see Fig.~\ref{fig:light-cone}A.

\subsection{Quasiparticle geometric picture}\label{sec:geometric_picture}

In the quasiparticle pair picture \cite{Calabrese2005}, an initial state $|\Psi_0\ra$, being excited from the point of view of a quenched Hamiltonian~$\hat H$, serves as a source of quasiparticle pair excitations. Each pair emitted from the same point in space is entangled and contribute to the total entanglement between $A$ and $B$, if at a time $t$ a pair is shared between the two regions. Accordingly, the entanglement entropy in the quasiparticle pair picture reads
\be\label{pair_picture_entropy}
S(t) = 2t \int \limits_{\mathclap{2t |v(p)|<L}} \mathrm{d}p\, |v(p)|f(p)+L\int\limits_{\mathclap{2 t|v(p)| >L}} \mathrm{d}p\, f(p),
\ee
where $f(p)$ is a quasiparticle production rate and $v(p)=\mathrm{d}E(p)/\mathrm{d}p$ is a quasiparticle velocity. If there exists a maximal quasiparticle velocity $v_\mathrm{max}$, than the entanglement entropy grows always linearly and saturates as a consequence of finite subsystem size at times $t\gtrsim t^* =L/2$. The applicability of Eq.~\eqref{pair_picture_entropy} goes beyond a simple qualitative understanding. In fact, in the thermodynamic limit $t,L\rightarrow \infty, t/L=\mbox{const.}$ the predictions of the quasiparticle picture exactly reproduce the behavior of entanglement entropy in 1D translationally invariant integrable models, based on  the knowledge of the steady state and its excitations. \cite{Alba2017}. For
arbitrary highly excited initial states,  Eq.~\eqref{pair_picture_entropy} describes the
leading contribution to the entanglement entropy \cite{Bertini_2018}. In the Appendix~\ref{app:crit_times}, using the quasiparticle formula, Eq.~\eqref{pair_picture_entropy}, we give an analytic form of the entanglement entropy for the staggered initial state and the homogeneous quench, Eq.~\eqref{postquench_hamiltonian}. There, we show that the analytical formula is nonanalytical at $t^*=L/2$, where the second derivative of entanglement entropy is discontinuous and we illustrate  that even for relatively small $L$'s the quasiparticle picture  quite accurately reproduces the numerical data, although the nonanalyticity can only be observed in the infinite subsystem limit. In the Appendix~\ref{app:crit_times} we also show that on contrary to the homogeneous case, where the entanglement entropy converges in a scaling limit $t,L \rightarrow \infty$, $t/L=const$, the saturation effects of entanglement entropy after a Rindler quench are far from being universal and the same scaling does not apply. 
For this reason, throughout the manuscript we focus on the entanglement entropy behavior up to times before saturation effects take place, i.e. for times smaller then the crossover time $t^*$.

\section{Correlation spreading}\label{sec:correlations}

The quasiparticle picture has a simple geometric interpretation. The 
 fastest quasiparticles propagate at the
 maximum velocity $v_{\mathrm max}=\max_k|\mathrm{d}E/\mathrm{d}k| =1$, the Lieb-Robinson velocity \cite{lieb1972finite}, which is the maximal velocity of correlation spreading  
\be\label{correlation_funtion}
C_{n,n'}(t) = \la \hat c^\dagger_n(t) \hat c_{n'} \ra\equiv \la \Psi_0 |\hat c^\dagger_n(t) \hat c_{n'} | \Psi_0 \ra,
\ee
where $|\Psi_0\ra$ is an initial state. In homogeneous systems, the correlations spread within linear light cones, which agrees with a quasiparticle pair picture. Geometrically, the contribution to the entanglement entropy $S(t)$ coming from each boundary between $A$ and $B$ is proportional to the length of an interval, $d(t)$,  covered by a light-cone  placed at this boundary. Assuming that the entanglement entropy is proportional to the number of degrees of freedom which can become correlated at a time $t< t^*=L/2$, the total entanglement entropy is then given by;
 \be\label{EE_pair-picture}
 S(t)\propto  2 d(t)  = 4  t.
 \ee
 
 The quasiparticle picture of counterpropagating pairs with opposite quasimomenta cannot be applied directly if the post-quench Hamiltonian does not possess translational invariance.  However, similarly to the homogeneous quench, we can relate the entanglement entropy growth with the spreading of correlations. In Fig.~\ref{fig:light-cone}(B) we plot the correlation function, Eq.~\eqref{correlation_funtion} for a~Rindler quench and for different choices of lattice positions $n'$. The exact numerical calculations show that the correlations spread nonlinearly within bent light-cones that are described by Eq.~\eqref{geodesics} with a very good agreement. In Fig.~\ref{fig:light-cone}(C) we present a cartoon picture of a bent light cone originating at a lattice site $m$. Through a simple calculation a distance, $d^-(t)$ and $ d^+(t)$, covered by a left and right part of the light cone accordingly is simply 
 \be
 d^\pm(t)=\mp m\left[1 -e^{\pm t}\right].
 \ee
 Now, we aim to apply semiclassical arguments to describe the entanglement dynamics. While this captures qualitative aspects of the entanglement dynamics, quantitative differences remain as we discuss in detail in the following. The direct application of a homogeneous result would lead to a superlinear growth of the entanglement entropy. Let us  heuristically assume the entanglement entropy is proportional to the number of \emph{distinct} pairs that can be correlated through each boundary between $A$ and $B$. This number is equal to $\min(d^+,d^-)=d^-$, see Fig.~\ref{fig:light-cone}(C), and the total entanglement entropy is
\be\label{EE_geometric-picture}
S( \tau) \propto 2 \left[ d_1^-(\tau) +  d_2^-(\tau) \right] = 4 \bar m \left[1- e^{-\tau/\bar m} \right],
\ee
where $\bar m= (m+L/2)$ is a position of the middle of $A$, and where $\tau = t  \, \bar{m}$ is a proper time of an observer placed at~$\bar m$. By switching to the proper time $\tau$, any stationary observer perceives the same local value of the speed of light $c(\bar m )/\bar{ m}=1$, which allows us to directly compare results with different observers' positions and different spacetimes.

The heuristic formula, Eq.~\eqref{EE_geometric-picture}, gives as an important insight into the behavior of entanglement entropy. First of all, it depends only on one parameter $\bar m$, and the dimensionless rescaled quantity $S(\tau \bar m)/\bar m$ is parameter-free. In Sec.~\ref{sec:numerics} we find numerically the same scaling for free lattice fermions. Furthermore, the formula predicts universal sublinear growth growth of entanglement. Let us look at two opposing limits: 
\begin{enumerate}
 \item if  $\tau/\bar m \ll 1 $, then $S(\tau)\propto 4 \tau$, so that locally and for early times a linear growth is recovered,
 \item if  $\tau/\bar m \gg 1 $, then $S(\tau) \propto \mbox{const.}$ .
\end{enumerate}
Let us stress that the second limit is not the finite-size effect, but rather stems from general properties of the Rindler spacetime where the horizon plays a role of an effective spacetime boundary.  

In the following sections we show that the entanglement generated by the Rindler quench has both of the above properties. Nevertheless, we also find that the heuristic formula, Eq.~\eqref{EE_geometric-picture}, is too naive and does not recover quantitatively the numerical data. In fact, Eq.~\eqref{EE_geometric-picture} considerably underestimate the entanglement entropy. One possible explanation is that the entanglement entropy requires important corrections from multiparticle entanglement, which is not captured in the above picture.

\section{Numerical results}\label{sec:numerics}

In this section we present numerical data for the entanglement entropy dynamics for the Rindler quench and stress the differences with the homogeneous case. We calculate the entanglement entropy by exact diagonalization of equal-time correlation matrices, see Sec.~\ref{sec:peschel_method}.  In Sec.~\ref{sec:initial_growth}  we show that the entanglement growth features universal behavior by achieving a data collapse, while in Sec.~\ref{sec:finite_size} we discuss finite-size effects in more details. Surprisingly,  the finite-size analysis allows us to identify a universal long-time asymptotic behavior.

\begin{figure}[bt]
 \includegraphics[width=1\columnwidth]{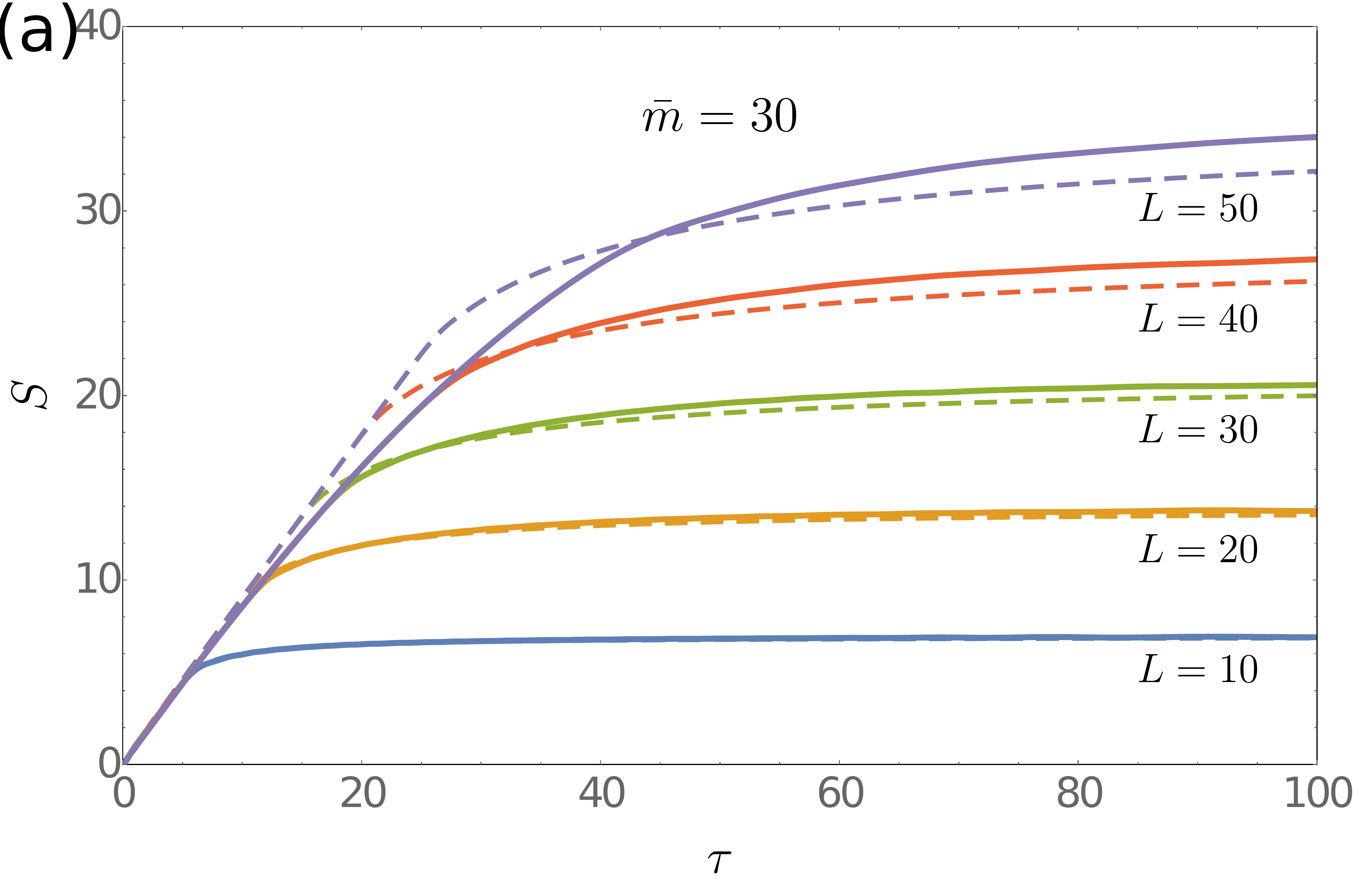}  
 \includegraphics[width=1\columnwidth]{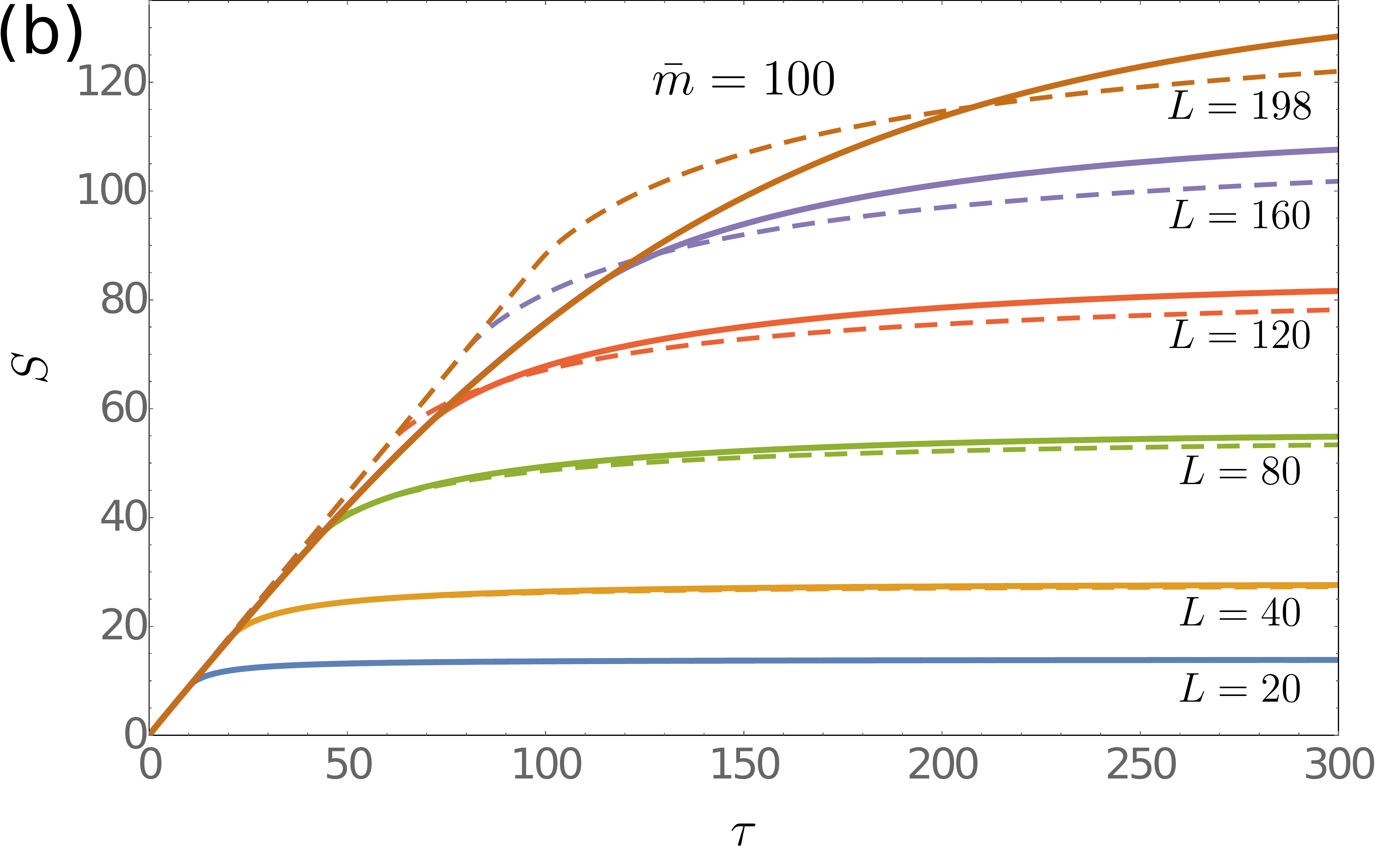}  
\caption{
The entanglement entropy obtained from the numerical computations. In order to extract a generic behaviour we fix the mean position of the subsystem  and perform finite size overlap of the data. Here we choose $\bar m = 30$ (panel (a)) and $\bar m =100 $ (panel (b)). For comparison we include data 
for both the Rindler (solid lines) and homogeneous quench (dashed lines). $\tau$ denotes a proper time of an observer, see discussion in the main text.
Apart from the finite subsystem size saturation to the $S_{max}=L\ln 2$, we find that all curves for different subsizes $L$'s and fixed mean positions $\bar m $ overlap, which allows us to identify a generic functional behavior.  
The numerical data confirm that $\bar m$ is the only relevant parameter and that the entanglement entropy grows sublinearly. Furthermore, we observe peculiar finite-size features of the Rindler quench: although the entanglement entropy grows slower then in the homogeneous case, the entanglement entropy saturates much faster.  }
\label{fig:ee_numerical_data}   
\end{figure} 

\begin{figure}[tb]
\includegraphics[width=1\columnwidth]{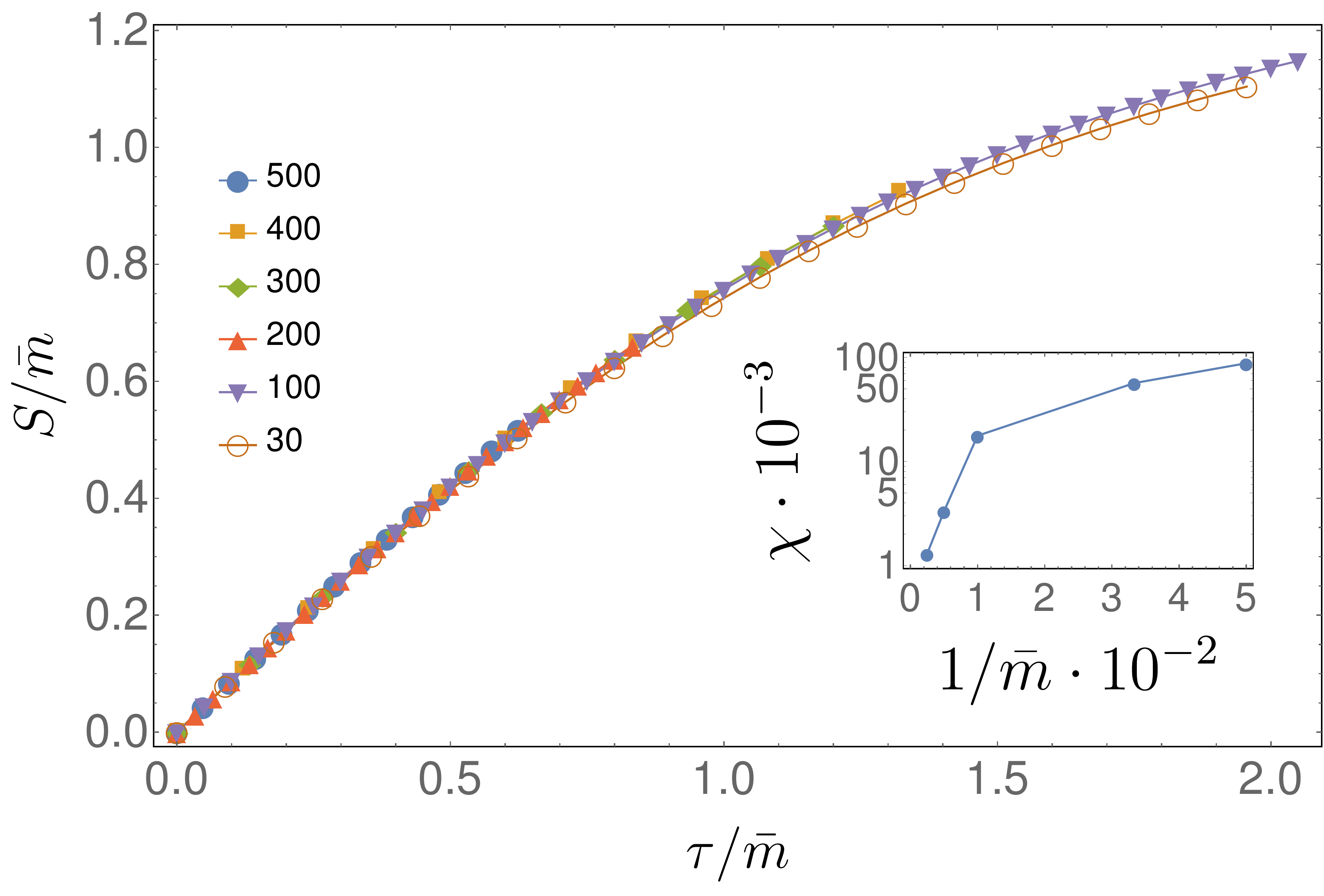}   
\caption{
Comparison of the entanglement entropy curves in the case of the Rindler quench, obtained from the finite-size data collapse for a fixed $\bar m=30,..,500$. All curves are rescaled accordingly to $s_{\bar m}(\tau')=S_{\bar m}(\bar m \tau')/\bar m$. The inset shows $\chi_{\bar m'}=[\int \mbox{d}\tau (s_{\bar m =500}(\tau)-s_{\bar m'}(\tau) )^2]^{1/2}$, which is the measure of distance between two functions. In the limit $ 1/\bar m  \rightarrow 0 $  we observe a converge of data.
}
\label{fig:ee_scaling}
\end{figure} 

\subsection{Universal sublinear growth}\label{sec:initial_growth}

In the case of the homogeneous quench, the entanglement entropy depends only on the subsystem size $L$ and the generic behavior can be numerically extracted by overlaping data for different subsystem sizes $L$. 
For the Rindler quench, due to the inhomogeneous metric, we find that the entanglement entropy depends also on the mean position of the subsystem $A$, see the semiclassical discussion in Sec.~\ref{sec:correlations}.  Therefore, in order to extract the universal behavior from the numerics, we need to fix a mean position $\bar m$ first and then perform the finite data overlap.  In Fig.~\ref{fig:ee_numerical_data} the entanglement entropy is plotted versus a proper time $\tau=t\, \bar m $ of observer located at $\bar m$.  For the sake of comparison, we include data for both 
 the Rindler (solid lines) and homogeneous (dashed lines) quenches. 

At early times and small subsystem sizes $L$ the data for Rindler and homogeneous quenches are indistinguishable
(this is expected as initial light cone bending from the point of view of local observers is small , see Eq.~\eqref{eq:geodesics_rindler}). The inhomogeneity of the spacetime becomes important at times scales $\tau \approx \bar m$, where the entanglement entropy for the Rindler quench exhibits sublinear growth behavior. For both quenches, we observe the collapse of the data and a generic entanglement function $S_{\bar m}(\tau)$ can be identified as an envelope of a family of curves, $ S_{\bar m,L}(\tau)$, with fixed $\bar m$ and different $L$.  In other words, $S_{\bar m} (\tau)$ can be defined a limit $S_{\bar m} (\tau) = \lim_{L\rightarrow  2\bar m} S_{\bar m, L } (\tau)  $.  

Let us now rescale the numerical curves $S_{\bar m}(\tau)$ for different values of $\bar m$ 
\be\label{universal_EE}
s_{\bar m}(\tau')= S_{\bar m}(\bar m \tau')/\bar m.
\ee

According to the semiclassical formula, Eq.~\eqref{EE_geometric-picture}, the $s_{\bar m}(\tau')$ curves should overlap. Indeed, although small deviations are found for small $\bar m$'s, in the limit $ 1/\bar m  \rightarrow 0 $  we observe a convergence of the data, i.e., $ s(\tau')$ defined as
 \be\label{ee_limit}
 s(\tau') = \lim_{1/\bar m  \rightarrow 0} s_{\bar m}(\tau')
 \ee
has a thermodynamic limit, see  Fig.~\ref{fig:ee_scaling}. 
This suggests a universal sublinear behavior of the entanglement entropy growth. On the contrary to the homogeneous case, the entropy production rate (the first derivative of entanglement entropy) decreases over time and,  as we argue in the next section, the entropy is asymptotically constant in a long-time limit.

\subsection{Asymptotic long-time dynamics}\label{sec:finite_size}

\begin{figure}[bt]
  \includegraphics[width=1\columnwidth]{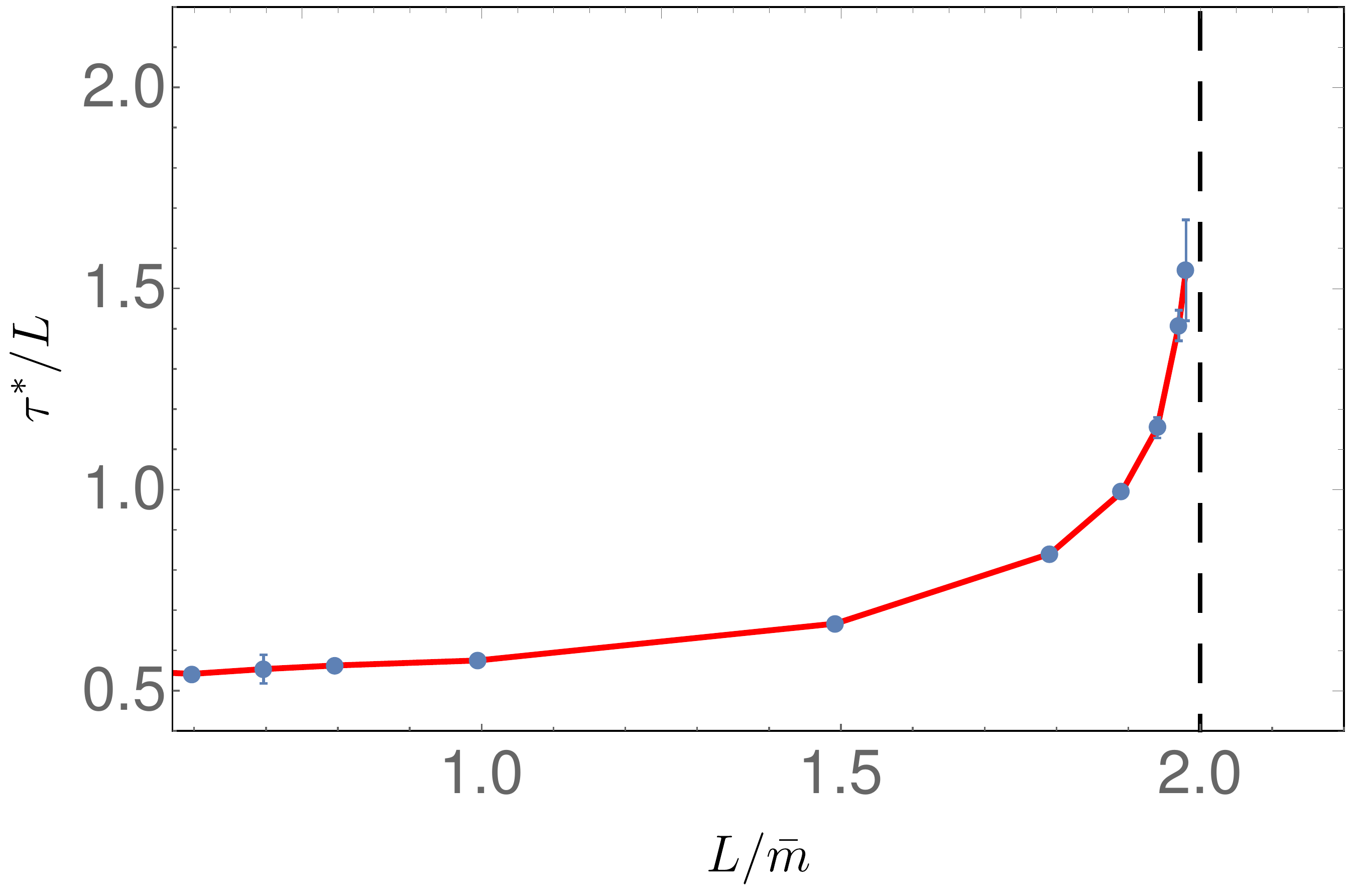}  
\caption{
Crossover times $\tau^*$ versus the ratio $L/\bar m $ obtained from numerical data. We definite a crossover time  $\tau^*$ of the entanglement entropy growth when the finite subsystem size saturation effects start to take place. Initially, for small ratios $L/\bar m \lessapprox 1$, the critical time is $\tau^* \approx L/2$, which agrees with the result from the homogeneous quench. The critical time grows drastically when the ratio ratio $L/\bar m $ increases and approaches its maximally attainable value equal to  $2 -2/\bar m$.   $\tau^*/L$ becomes infinite in the limit $\bar m \rightarrow \infty $ such that $L/\bar m \rightarrow 2$. 
}
\label{fig:critical_times}   
\end{figure} 

In this section we analyze finite subsystem size effects in entanglement entropy evolution, which allows us to identify universal asymptotic long-time dynamics in the thermodynamic limit.  Already in Fig.~\ref{fig:ee_numerical_data} one could notice two striking saturation effect differences between the Rindler and homogeneous quenches. First of all, a crossover time $\tau^*$~--~a time when the evolution starts to thermalize due to a finite size of a subsystem~--~increases together the growth of a subsystem size $L$.  Secondly, the finite size thermalization is much faster. 

Let us first quantify the first observation. We recall that $\bar m $ denotes a middle position of a subsystem $A$ of length $L$, and therefore the  $L/\bar m \in [1/\bar m , 2$-$2/\bar m$]. The lower limit corresponds to a situation when $A$ of a unit length is far away from the origin $m=1$, while the upper limit corresponds to $A$ located maximally close to the origin. Previously, we have identified a generic function $S_{\bar m}(\tau)$ which is an envelope of a family of curves $S_{\bar m,L}(\tau)$ with fixed $L$. We define a crossover time $\tau^*(L)$ as $|S_{\bar m,L}(\tau)-S_{\bar m}(\tau)| < \epsilon$   and on  Fig.~\ref{fig:critical_times}  we plot $\tau^*/L$ as function of $L/\bar m$   for $\bar m = 200$ and $\epsilon =0.03$. While the choice for the threshold $\epsilon$ is, of course, arbitrary, we find that the final result for $\tau^*(L)$ doesn't depend crucially on it as long as $\epsilon$ is sufficiently small. As expected, we observe that $\tau ^* \approx L/2$ when the subsystem $A$ is small comparing to its distance to the origin, which recovers the known homogeneous result. This is perfectly understandable, since for small $L/m$ the effective speed of light $c(m)\propto m$ is locally constant and does not change significantly on the extent of a subsystem, i.e., 
\be
c(m+L/2)/c(m-L/2) \approx 1+L/m.
\ee
On the other hand,  $\tau^*$ increases drastically when $L/\bar m$ ratio approaches its maximal value $2-2/\bar m$ and in the thermodynamic limit $\bar m \rightarrow \infty $, the function   $\tau^*/L$ becomes infinite  at $L/\bar m =2 $, see Fig.~\ref{fig:critical_times}. From this seemingly small result we can actually infer the asymptotic long time behavior of the universal curve $s(\tau)$, Eq.~\eqref{ee_limit}. Since $ \forall_{\tau} S_{\bar m, L } (\tau) \leq L \ln 2$ and $S_{\bar m} (\tau) = \lim_{L\rightarrow  2\bar m} S_{\bar m, L } (\tau)  $ , then  
\be
\lim_{L\rightarrow  2\bar m} S_{\bar m, L } (\tau^*)   = S_{\bar m } (\infty) = 2 \bar m \ln 2
\ee
and consequently $s(\tau)$  has a horizontal asymptote 
\be\label{ee_asymptote}
\lim_{\tau \rightarrow \infty}s(\tau)  =  2 \, \ln 2. 
\ee
Let us stress that although the functional behavior of the entanglement entropy might be different for different initial states, the limit in Eq.\eqref{ee_asymptote}  is generic for a wide class of highly excited homogeneous initial states. In physical terms, the thermodynamic limit $\bar m \rightarrow \infty $ while $L/\bar m\rightarrow 2 $ means that one edge of the subsystem $A$ is placed basically at the horizon, where the local light velocity vanishes asymptotically. As a consequence, it would require an infinite time for the signal to propagate throughout the subsystem and the crossover time $\tau^*$ has to diverge accordingly. Note that this behavior was already predicted on a semiclassical level, Sec.~\ref{sec:correlations}. 

\begin{figure}[bt]
  \includegraphics[width=1\columnwidth]{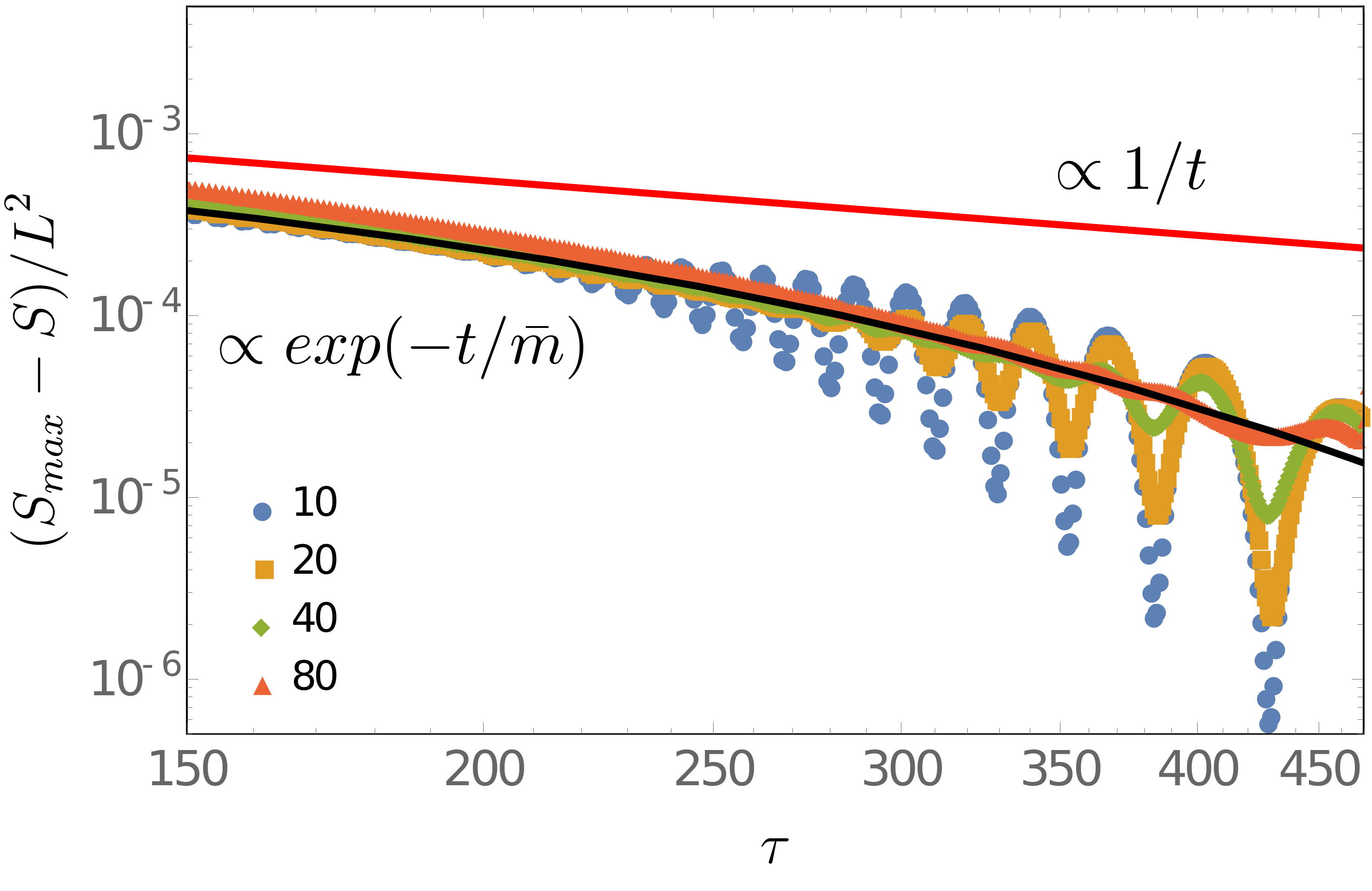}  
\caption{The distance to the asymptotic maximal value of entanglement entropy $|S(\tau)-S_{max}|$ versus time for a fixed value $\bar m =100$. Although initially the entanglement entropy for a Rindler quench grows sublinearly, we find that it approaches the thermodynamic value $S_{\textrm{max}} = L \ln 2$ exponentially fast, while for a global quench the entanglement entropy (red curve) saturates as $S \propto L^2/t$.
}
\label{fig:saturation}   
\end{figure}

Finally, we investigate the entanglement entropy saturation rate, i.e., how fast the subsystem thermalizes at times scales much larger than the crossover time. 
In Fig.~\ref{fig:saturation} plot the distance $|S(\tau)-S_{\textrm{max}}|$ to the maximal value $S_{\textrm{max}}=L \ln 2$ in time. It is known that in a case of global quench the entanglement entropy saturates as $L^2/t$ \cite{Calabrese2005} (see also Appendix~\ref{app:crit_times}). On the other hand, in a case of a Rindler quench, we can see that after times $\tau\approx\bar m$ a subsystem thermalizes exponentially fast, which we find very intriguing: although the entanglement entropy grows sublinearly, it saturates much faster than in a homogeneous case. 
Last but not least let us stress that the saturation effects of the entanglement entropy for a Rindler quench are in general not universal and the scaling limit  $t,L \rightarrow \infty$, $t/L = const$  is a proper one only when the crossover time diverges. In this case we can quantify the numerical results by fitting the prediction from the unitary circuits analysis, see Sec.~\ref{sec:randomcircuits}. There, we obtain that $S(t) \propto\tanh (const. \, t/\bar m) =  \tanh (2 \, const.\, t/l)$.

\section{Random Unitary Circuits}\label{sec:randomcircuits}

In the previous sections we have investigated the entanglement entropy evolution after the Rindler quench in a simple system of free fermions on a lattice. In particular, we have found that a simple heuristic semiclassical arguments (Sec.~\ref{sec:correlations}) give us important insight into the universal behavior of entanglement dynamics, although it fails to quantitatively reproduce the numerical data (Sec.~\ref{sec:numerics}). The semiclassical arguments cannot account for  multiparticle contribution to the entanglement, which is most probably the reason of quantitatively differences.
On the other hand, such mutliparticle contribution can be captured in random unitary circuit setups, which provide  minimally structured toy models of generic non-integrable systems.  For this reason, in this part we estimate the  entanglement entropy for a specific system of random unitary circuits, where the coarse grained entanglement dynamics is equivalent to the minimal membrane description \cite{jonay2018coarsegrained}. 

\begin{figure}[bt]
  
\includegraphics[width=1\columnwidth]{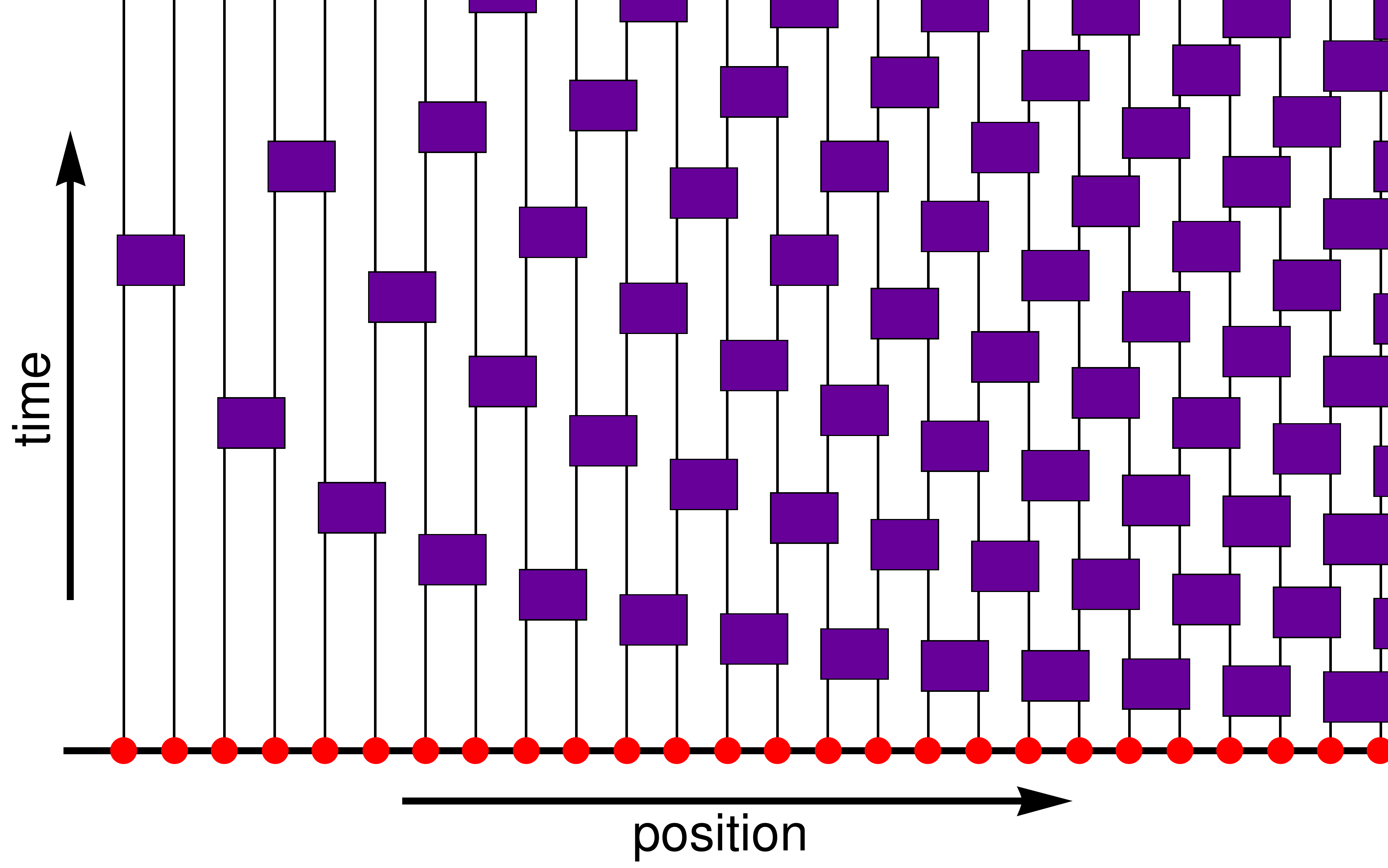}  
\caption{ A schematic structure for simulating a random unitary circuits with a Rindler metric. Red points represents an initial spin configuration and purple rectangles gates are random unitary gates. The rate at which random unitary gates are applied is an inverse function of a distance to the horizon placed at the left boundary of the system.
}
\label{fig:rand_unitaries}   
\end{figure} 

Let us denote $S(x,t)$ for the entanglement entropy for an arbitrary quantum state, where $x$ is a position of a bipartite cut. If, like in the previous section, we would rather consider a bipartition where a subsystem  $A$ is a finite segment $A=[m_1,m_2]$, then  the entanglement entropy of such a bipartition is  $S_A(t) = S(m_1,t) +S(m_2,t)$.
After \cite{jonay2018coarsegrained} we write down the equation for the leading order coarse-grained dynamics of the local rate of the entanglement entropy
\be\label{eq1}
\partial_{t} S(x,t) = \Gamma [s= \partial_x S (x,t)],
\ee
where $\Gamma(s)$ is a production rate
\be
\Gamma (s) = \gamma (1- \alpha s^2),
\ee
where $\alpha$, $\gamma$ are free parameters. While Eq.\eqref{eq1} is universal for a generic random unitary circuit setup, a specific model can be achieved by fixing free parameters $\alpha$ and $\gamma$, see Ref.~\cite{jonay2018coarsegrained}.  Let us note that in general the above formula should be treated as Taylor expansion of the entanglement line tension, where the higher order terms do not contribute to the coarse grained dynamics as long as $\partial_x S (x,t)$ is sufficiently small. 

Our goal is to describe the entanglement entropy dynamics in a Rindler quench scenario.  To make it work, we simply notice that in Rindler Universe, Eq.~\eqref{rindler_metric}, due to spatially varying speed of light $c(x)=x$, the local dynamics is the slower the closer are to $x=0$, and so $\gamma=\beta x $ should be proportional to $x$, yielding

\be
\partial_t S(x,t) = x \beta \left[ 1- \alpha (\partial_x S(x,t))^2 \right]. 
\ee
In a same way, to simulate a random circuits with a Rindler metric one should apply random unitary gates with rates proportional to the inverse distance to the horizon, see Fig.~\ref{fig:rand_unitaries} for an illustration. 
Now, by utilizing an ansatz for separation of variables
\be
S(x,t)= x f(t) 
\ee
it is straightforward to solve the corresponding differential equation 
\be
\partial_t f(t) = \beta (1-\alpha f(t)^2),
\ee
which has an elementary solution
\be
f(t) = \frac{1}{\sqrt{\alpha}} \tanh(\sqrt{\alpha} \beta t).
\ee
Finally, then the entanglement entropy between $A$ and its complement reads 
\be\label{entropy_random_unitaries}
S_A(\tau) = \frac{2\bar m}{\sqrt{\alpha}} \tanh (\sqrt{\alpha} \beta \tau /\bar  m),
\ee
where $\bar m =(m_1+m_2)/2$ and $\tau = \bar m t$ is a proper time of an observer located at $\bar m$.

\begin{figure}[bt]
 \includegraphics[width=1\columnwidth]{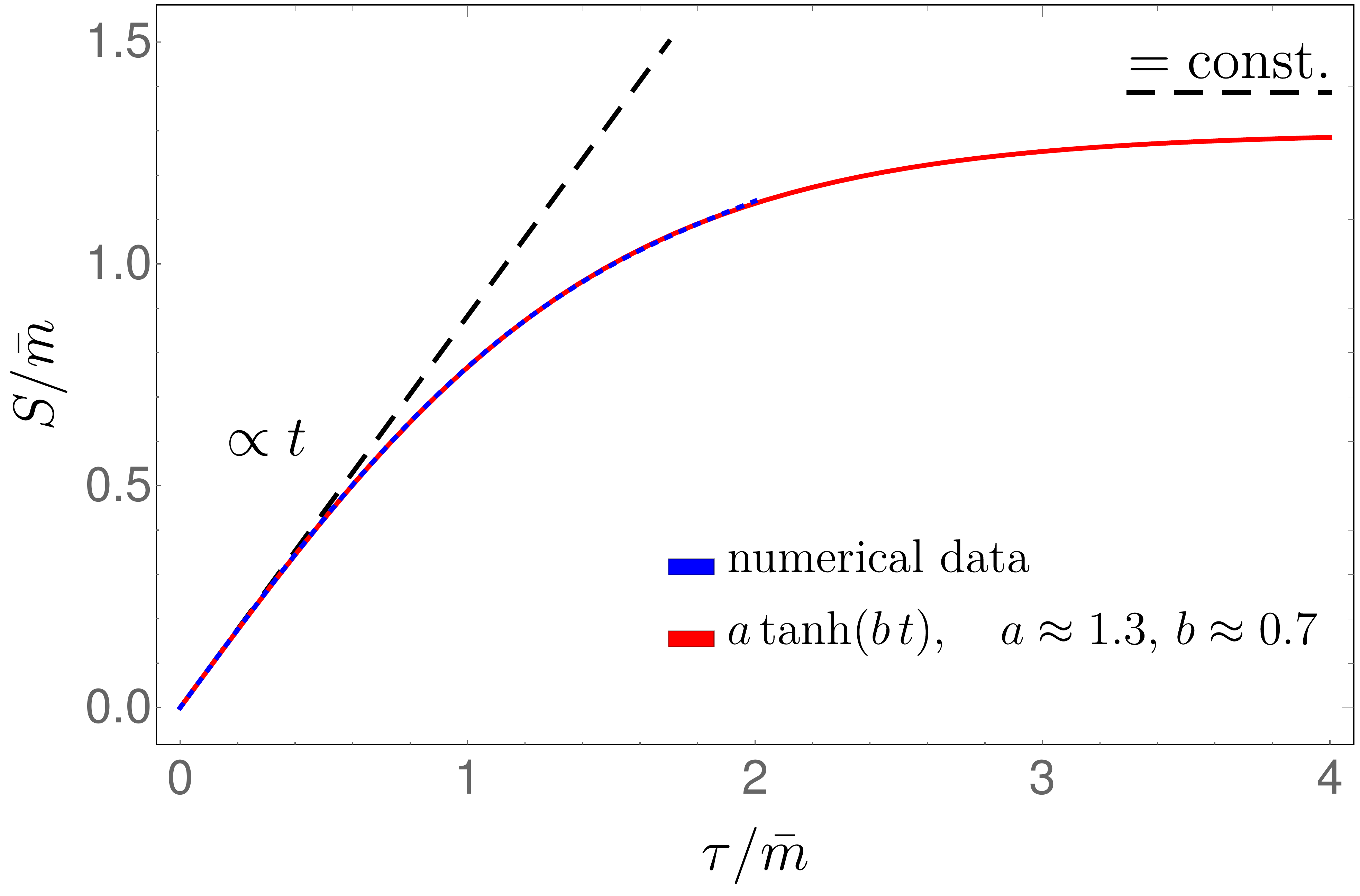}  
\caption{
Comparison of the entanglement entropies for a Rindler quench and for a free fermionic chain obtained from exact numerical data  (blue dashed) and from  the random unitary circuits model (red).
}
\label{fig:ee_comparison}
\end{figure} 

As we have obtained from the semiclassical analysis, where $S(\tau) = \bar{m}(1-e^{-\tau/\bar m} )$, see Sec.~\ref{sec:correlations}, the entanglement entropy growth predicted by the random unitary circuit, Eq.~\eqref{entropy_random_unitaries}, is linear at early times and attains a constant values at long times scales.
On the contrary, we know that the entanglement entropy of random unitary circuit models could in principle capture multiparticle contribution to the entanglement. Therefore, although the two formulas obey the same scaling $s(\tau)=  S(\tau \bar m)/\bar{m}$ and are qualitatively similar, we can expect quantitative differences both in the short and long time behavior. In particular, at short time scales the first non vanishing correction to the linear growth in $S(\tau)$ is quadratic while in $S_A(\tau)$ the first correction  is cubic.  At long time scales it is straightforward to find that $S_A(\tau) \propto 1 - 2 e^{-2 \sqrt{\alpha}\beta \tau/\bar m} $.

Notice that the entanglement entropy $S_A(\tau)$, Eq.~\eqref{entropy_random_unitaries}, depends on two free parameters which correspond to  different choices of the entropy production rate.  Since $\alpha,\beta$ are two unknown model specific parameters, $S_A(\tau)$ should also recover the numerical results obtained for non-interacting fermions. Let us therefore fit $s_A(\tau)=  S_A(\tau \bar m)/\bar{m}$ to the numerical data from the Sec.~\ref{sec:numerics}. We plot the results in Fig.~\ref{fig:ee_comparison} and find that the curve remarkably agrees with the numerical data with fitting coefficients
$
\alpha\approx 2.4, \beta \approx 0.45
$.  This suggests that our result captures universal properties of entanglement entropy dynamics after the Rindler quench.

\section{Conclusions}\label{sec:conclusions}

The main goal of this work was to study entanglement growth in inhomogeneous spacetimes, where correlations do not propagate within  straight light cones.  We have taken as an example the (1+1) Rindler metric, which is known to have a spacetime horizon that strongly distorts the light cones in its vicinity.  We have shown that the entanglement initially grows sublinearly and in a long-time limit is asymptotically constant. This behavior can be qualitatively understood via semiclassical arguments that base on the knowledge of correlation spreading. Correlations spread within distorted light-cones that are described by null geodesics of the Rindler spacetime. This suggests that we have captured basic properties of the continuum theory and that the similar reasoning can be applied to other spacetimes. 

We have found indications that our observations are universal. For the paradigmatic example of an ergodic system, we have studied the entanglement growth also for a random unitary circuit model. Choosing a specific random unitary circuit setup, we have derived a leading coarse-grained dynamics of the  entanglement  entropy that has the same characteristics as in a case of free fermions on the lattice.  This suggests that our results are applicable to a generic quantum system. This opens the way towards studying entanglement production in more general inhomogeneous spacetimes.

\section*{ACKNOWLEDGMENTS}

We acknowledge a very fruitful discussion with Adam Nahum. A.K. would like to thank Alessio Celi, Maciej Lewenstein and Emanuele Tirrito for previous discussions concerning Rindler physics. A.K. acknowledge the support of the Foundation for Polish Science (FNP) and the support of the National Science Centre, Poland via Projects No.~2016/21/B/ST2/01086 (A.K.). This project has received funding from the European Research Council (ERC) under the European Unions Horizon 2020 research and innovation programme (grant agreement No. 853443), and M.~H. further acknowledges support by the Deutsche Forschungsgemeinschaft via the Gottfried Wilhelm Leibniz Prize program.

\begin{appendix}
 
\section{Nonanalytic behavior of the entanglement entropy}\label{app:crit_times}

\begin{figure}[t]
  \includegraphics[width=1\columnwidth]{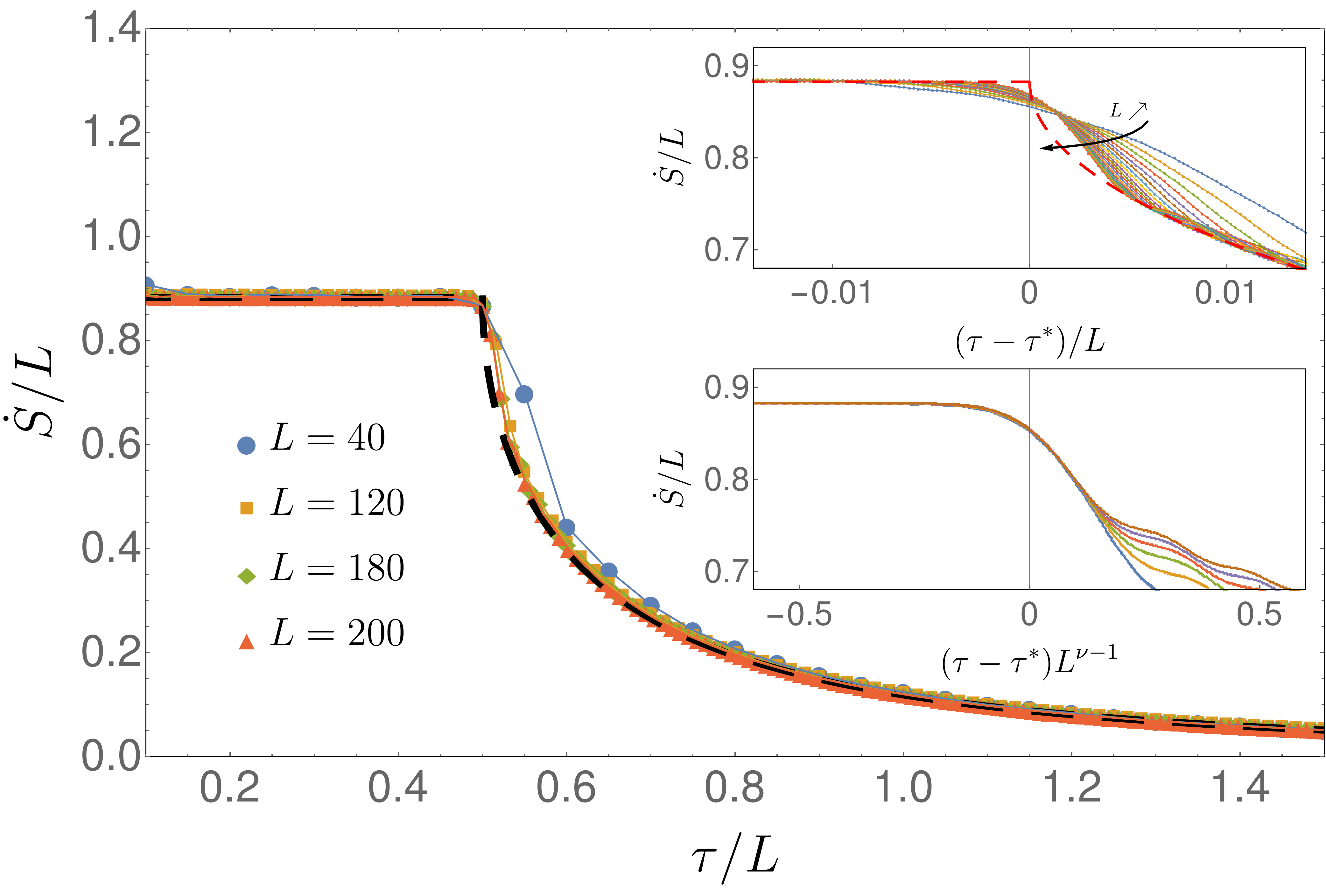}  
\caption{The Entanglement entropy after a global quench.  The analytical prediction for the entanglement entropy $\dot s(t)$ (black dashed line) and the exact numerical data for finite subsystem sizes. The upper inset shows that increasing the subsystem size (up to $L=2\cdot10^4$), the numerical data get closer to the analytical prediction. The lower inset illustrates a collapse of data after a finite-size scaling. The scaling coefficient is $\nu \approx 0.5$.  
}
\label{fig:global_quench:critical_time}   
\end{figure} 

The entanglement entropy due to quasiparticle picture \cite{Calabrese2005} reads 
\be\label{pair_picture_entropy_appendix}
S(t) = 2t \int \limits_{\mathclap{2t |v(p)|<L}} \mathrm{d}p\, |v(p)|f(p)+L\int\limits_{\mathclap{2 t|v(p)| >L}} \mathrm{d}p\, f(p),
\ee
where $f(p)$ is a quasiparticle production rate, that we obtain from the properties of the stationary state \cite{Bertini_2018} , and $v(p)=\mathrm{d}E(p)/\mathrm{d}p =  \cos(p)$ is a simple lattice dispersion relation for free particles. For a global homogeneous quench that we consider the main part of the article we obtain

\be \label{ee_analitic}
S(t) = 
\left\{
\begin{matrix}
S_{\rm lin}(t) & , &
\ 0 \leq t \leq L/2&\\
S_{\rm sat}(t) &,  & t>L/2&
\end{matrix}
\right. ,
\ee
where
\be
S_{\rm lin}(t) =\frac{4  \ln 2}{\pi} t,
\ee
\be
S_{\rm sat}(t)= \frac{2 L \ln 2 }{\pi}\left[ \arccos\left(\frac{L}{2 t}\right)  +  \frac{2t-\sqrt{4 t^2-L^2}}{L}  \right].
\ee

From general considerations, Calabrese and Cardy \cite{Calabrese2005} argue that at late times $t\ll L/2$ the entanglement entropy $S(t)$ saturates as $\propto L^2/t$. In our specific case, we can calculate the asymptotic behaviour exactly. The Taylor expansion of $S_{\textrm sat} (t) $ at small $x=L/2t$ yields
\be
S_{\mathrm sat} (t \ll L/2) =  L\ln 2 - \frac{L^2 \ln 2}{2\pi t} + o\left((L/2t)^2\right).
\ee
It is convenient to rescale $s(t)\equiv S(t L)/L$, which does not depend on $L$, such that the thermodynamic limit $t,L\rightarrow \infty$, but $t/L$=const. is  straightforward.  The function $s(t)$ belongs to the $\mathcal{C}^1$ differentiability class, i.e., its derivative is not differentiable at $t=t^*$. We plot $s(t)$ together with the numerical data in Fig.~\ref{fig:global_quench:critical_time}. We see that even for relatively small $L$'s the numerical data reproduces the analytic curve almost perfectly, where the divergence can be observed in a vicinity of $t^*$. Yet, the finite-size analysis confirms non analytical  behaviour of $s(t)$ in the thermodynamic limit, see insets of Fig.~\ref{fig:global_quench:critical_time}.

\begin{figure}[tbh]
  \includegraphics[width=1\columnwidth]{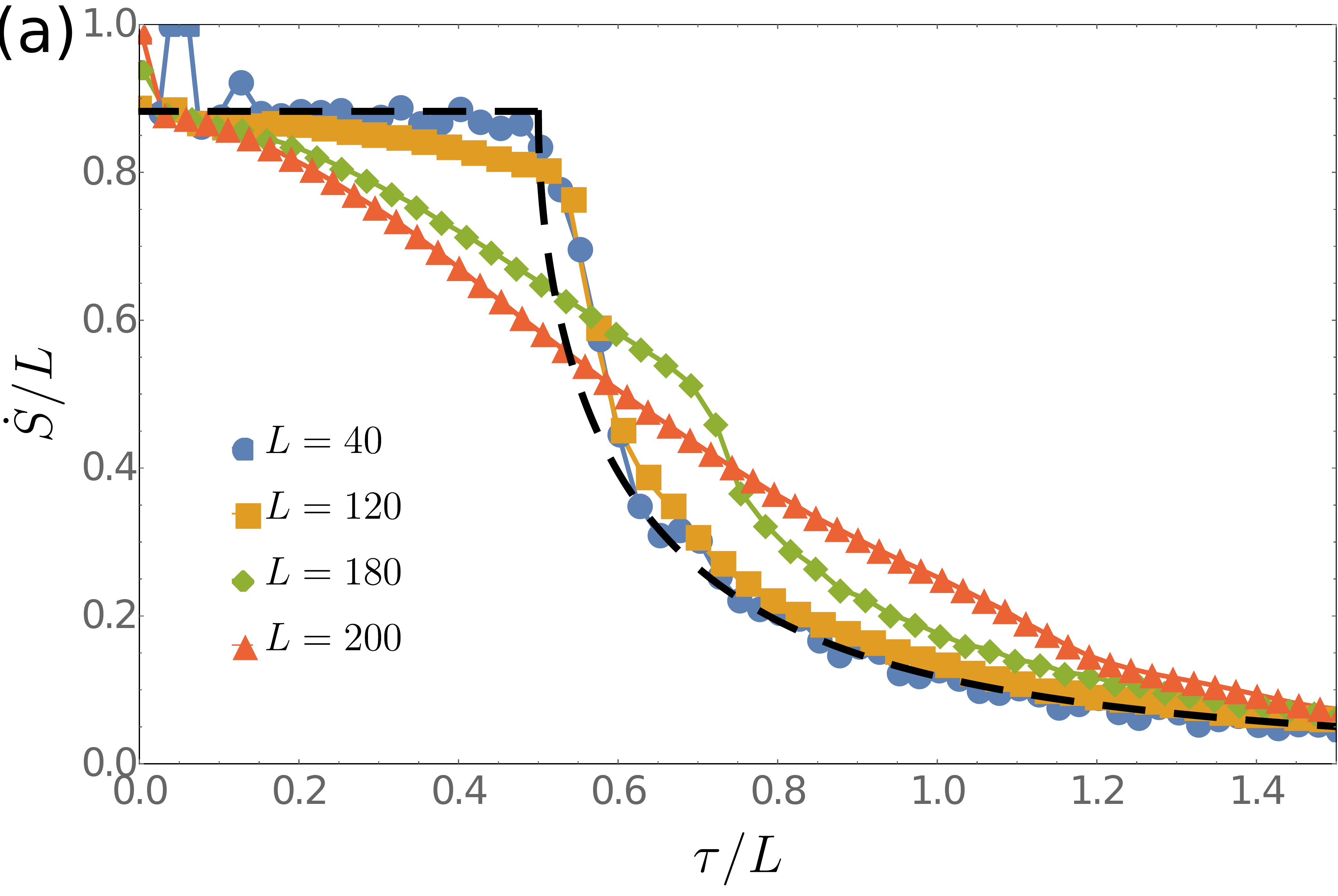}  
\includegraphics[width=1\columnwidth]{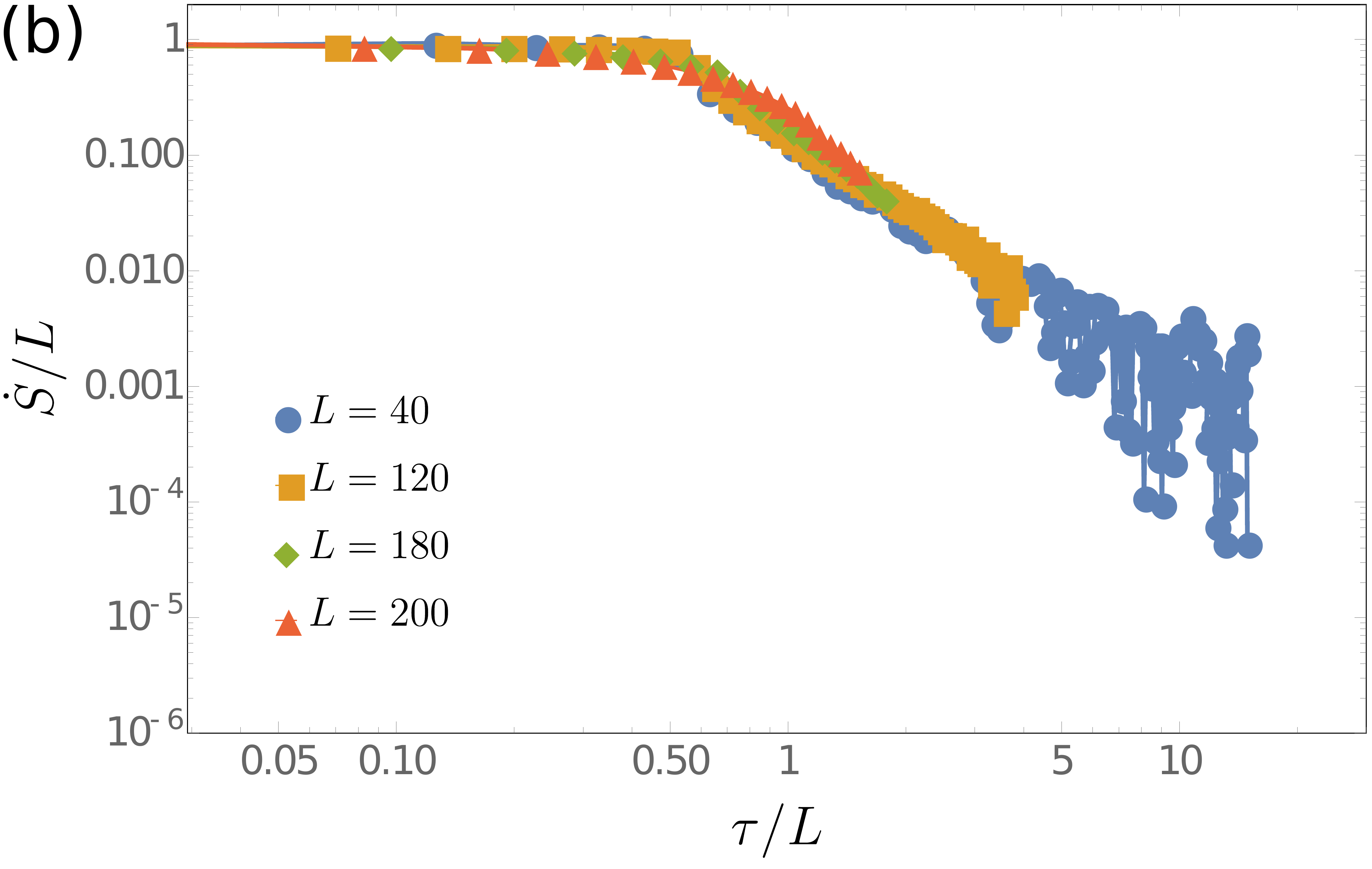} 
\caption{The entanglement entropy after a Rindler quench. The numerical derivative of the entanglement entropy. The black dashed line is an analytic line $\dot s(t)$ obtained for a global quench. We find that with the increasing $L$ the derivative becomes smoother around $t^*=L/2$.  Panels (a) and (b) shows the same data on linear and logarithmic scales accordingly. The computations  were  performed for $\bar m = 200$. 
}
\label{fig:rindler_quench:critical_time}   
\end{figure} 

In the main text of this article, we have shown that the saturation effects of entanglement entropy after a Rindler quench are far from being universal. In Fig.~\ref{fig:rindler_quench:critical_time}, as before, we plot the first derivative of the entanglement entropy for different subsystem sizes $L$ and fixed $\bar m =200$ , see Fig.~\ref{fig:global_quench:critical_time} for a comparison. The derivative gets smoother with increasing $L$ and the crossover point becomes less pronounced.

\end{appendix}


\begin{thebibliography}{71}
\expandafter\ifx\csname natexlab\endcsname\relax\def\natexlab#1{#1}\fi
\expandafter\ifx\csname bibnamefont\endcsname\relax
  \def\bibnamefont#1{#1}\fi
\expandafter\ifx\csname bibfnamefont\endcsname\relax
  \def\bibfnamefont#1{#1}\fi
\expandafter\ifx\csname citenamefont\endcsname\relax
  \def\citenamefont#1{#1}\fi
\expandafter\ifx\csname url\endcsname\relax
  \def\url#1{\texttt{#1}}\fi
\expandafter\ifx\csname urlprefix\endcsname\relax\def\urlprefix{URL }\fi
\providecommand{\bibinfo}[2]{#2}
\providecommand{\eprint}[2][]{\url{#2}}

\bibitem[{\citenamefont{Bengtsson and
  Zyczkowski}(2006)}]{bengtsson_zyczkowski_2006}
\bibinfo{author}{\bibfnamefont{I.}~\bibnamefont{Bengtsson}} \bibnamefont{and}
  \bibinfo{author}{\bibfnamefont{K.}~\bibnamefont{Zyczkowski}},
  \emph{\bibinfo{title}{Geometry of Quantum States: An Introduction to Quantum
  Entanglement}} (\bibinfo{publisher}{Cambridge University Press},
  \bibinfo{year}{2006}).

\bibitem[{\citenamefont{Horodecki et~al.}(2009)\citenamefont{Horodecki,
  Horodecki, Horodecki, and Horodecki}}]{Horodecki2009}
\bibinfo{author}{\bibfnamefont{R.}~\bibnamefont{Horodecki}},
  \bibinfo{author}{\bibfnamefont{P.}~\bibnamefont{Horodecki}},
  \bibinfo{author}{\bibfnamefont{M.}~\bibnamefont{Horodecki}},
  \bibnamefont{and}
  \bibinfo{author}{\bibfnamefont{K.}~\bibnamefont{Horodecki}},
  \bibinfo{journal}{Rev. Mod. Phys.} \textbf{\bibinfo{volume}{81}},
  \bibinfo{pages}{865} (\bibinfo{year}{2009}),
  \urlprefix\url{https://link.aps.org/doi/10.1103/RevModPhys.81.865}.

\bibitem[{\citenamefont{Ekert}(1991)}]{Ekert1991}
\bibinfo{author}{\bibfnamefont{A.~K.} \bibnamefont{Ekert}},
  \bibinfo{journal}{Phys. Rev. Lett.} \textbf{\bibinfo{volume}{67}},
  \bibinfo{pages}{661} (\bibinfo{year}{1991}),
  \urlprefix\url{https://link.aps.org/doi/10.1103/PhysRevLett.67.661}.

\bibitem[{\citenamefont{Gisin et~al.}(2002)\citenamefont{Gisin, Ribordy,
  Tittel, and Zbinden}}]{Gisin2002}
\bibinfo{author}{\bibfnamefont{N.}~\bibnamefont{Gisin}},
  \bibinfo{author}{\bibfnamefont{G.}~\bibnamefont{Ribordy}},
  \bibinfo{author}{\bibfnamefont{W.}~\bibnamefont{Tittel}}, \bibnamefont{and}
  \bibinfo{author}{\bibfnamefont{H.}~\bibnamefont{Zbinden}},
  \bibinfo{journal}{Rev. Mod. Phys.} \textbf{\bibinfo{volume}{74}},
  \bibinfo{pages}{145} (\bibinfo{year}{2002}),
  \urlprefix\url{https://link.aps.org/doi/10.1103/RevModPhys.74.145}.

\bibitem[{\citenamefont{Pirandola et~al.}(2019)\citenamefont{Pirandola,
  Andersen, Banchi, Berta, Bunandar, Colbeck, Englund, Gehring, Lupo, Ottaviani
  et~al.}}]{Pirandola2019}
\bibinfo{author}{\bibfnamefont{S.}~\bibnamefont{Pirandola}},
  \bibinfo{author}{\bibfnamefont{U.~L.} \bibnamefont{Andersen}},
  \bibinfo{author}{\bibfnamefont{L.}~\bibnamefont{Banchi}},
  \bibinfo{author}{\bibfnamefont{M.}~\bibnamefont{Berta}},
  \bibinfo{author}{\bibfnamefont{D.}~\bibnamefont{Bunandar}},
  \bibinfo{author}{\bibfnamefont{R.}~\bibnamefont{Colbeck}},
  \bibinfo{author}{\bibfnamefont{D.}~\bibnamefont{Englund}},
  \bibinfo{author}{\bibfnamefont{T.}~\bibnamefont{Gehring}},
  \bibinfo{author}{\bibfnamefont{C.}~\bibnamefont{Lupo}},
  \bibinfo{author}{\bibfnamefont{C.}~\bibnamefont{Ottaviani}},
  \bibnamefont{et~al.}, \emph{\bibinfo{title}{Advances in quantum
  cryptography}} (\bibinfo{year}{2019}), \eprint{1906.01645}.

\bibitem[{\citenamefont{Bennett et~al.}(1993)\citenamefont{Bennett, Brassard,
  Cr\'epeau, Jozsa, Peres, and Wootters}}]{Bennett1993}
\bibinfo{author}{\bibfnamefont{C.~H.} \bibnamefont{Bennett}},
  \bibinfo{author}{\bibfnamefont{G.}~\bibnamefont{Brassard}},
  \bibinfo{author}{\bibfnamefont{C.}~\bibnamefont{Cr\'epeau}},
  \bibinfo{author}{\bibfnamefont{R.}~\bibnamefont{Jozsa}},
  \bibinfo{author}{\bibfnamefont{A.}~\bibnamefont{Peres}}, \bibnamefont{and}
  \bibinfo{author}{\bibfnamefont{W.~K.} \bibnamefont{Wootters}},
  \bibinfo{journal}{Phys. Rev. Lett.} \textbf{\bibinfo{volume}{70}},
  \bibinfo{pages}{1895} (\bibinfo{year}{1993}),
  \urlprefix\url{https://link.aps.org/doi/10.1103/PhysRevLett.70.1895}.

\bibitem[{\citenamefont{Pirandola et~al.}(2015)\citenamefont{Pirandola, Eisert,
  Weedbrook, Furusawa, and Braunstein}}]{Pirandola2015}
\bibinfo{author}{\bibfnamefont{S.}~\bibnamefont{Pirandola}},
  \bibinfo{author}{\bibfnamefont{J.}~\bibnamefont{Eisert}},
  \bibinfo{author}{\bibfnamefont{C.}~\bibnamefont{Weedbrook}},
  \bibinfo{author}{\bibfnamefont{A.}~\bibnamefont{Furusawa}}, \bibnamefont{and}
  \bibinfo{author}{\bibfnamefont{S.~L.} \bibnamefont{Braunstein}},
  \bibinfo{journal}{Nature Photonics} \textbf{\bibinfo{volume}{9}},
  \bibinfo{pages}{641–652} (\bibinfo{year}{2015}), ISSN
  \bibinfo{issn}{1749-4893},
  \urlprefix\url{http://dx.doi.org/10.1038/nphoton.2015.154}.

\bibitem[{\citenamefont{Feynman}(1982)}]{Feynman1982}
\bibinfo{author}{\bibfnamefont{R.~P.} \bibnamefont{Feynman}},
  \bibinfo{journal}{International Journal of Theoretical Physics}
  \textbf{\bibinfo{volume}{21}}, \bibinfo{pages}{467} (\bibinfo{year}{1982}),
  \urlprefix\url{https://doi.org/10.1007/BF02650179}.

\bibitem[{\citenamefont{Steane}(1998)}]{Steane1998}
\bibinfo{author}{\bibfnamefont{A.}~\bibnamefont{Steane}},
  \bibinfo{journal}{Reports on Progress in Physics}
  \textbf{\bibinfo{volume}{61}}, \bibinfo{pages}{117} (\bibinfo{year}{1998}),
  \urlprefix\url{https://doi.org/10.1088%2F0034-4885%2F61%2F2%2F002}.

\bibitem[{\citenamefont{Ladd et~al.}(2010)\citenamefont{Ladd, Jelezko,
  Laflamme, Nakamura, Monroe, and O’Brien}}]{Ladd2010}
\bibinfo{author}{\bibfnamefont{T.~D.} \bibnamefont{Ladd}},
  \bibinfo{author}{\bibfnamefont{F.}~\bibnamefont{Jelezko}},
  \bibinfo{author}{\bibfnamefont{R.}~\bibnamefont{Laflamme}},
  \bibinfo{author}{\bibfnamefont{Y.}~\bibnamefont{Nakamura}},
  \bibinfo{author}{\bibfnamefont{C.}~\bibnamefont{Monroe}}, \bibnamefont{and}
  \bibinfo{author}{\bibfnamefont{J.~L.} \bibnamefont{O’Brien}},
  \bibinfo{journal}{Nature} \textbf{\bibinfo{volume}{464}},
  \bibinfo{pages}{45–53} (\bibinfo{year}{2010}), ISSN
  \bibinfo{issn}{1476-4687},
  \urlprefix\url{http://dx.doi.org/10.1038/nature08812}.

\bibitem[{\citenamefont{Hauke et~al.}(2012)\citenamefont{Hauke, Cucchietti,
  Tagliacozzo, Deutsch, and Lewenstein}}]{Hauke2012}
\bibinfo{author}{\bibfnamefont{P.}~\bibnamefont{Hauke}},
  \bibinfo{author}{\bibfnamefont{F.~M.} \bibnamefont{Cucchietti}},
  \bibinfo{author}{\bibfnamefont{L.}~\bibnamefont{Tagliacozzo}},
  \bibinfo{author}{\bibfnamefont{I.}~\bibnamefont{Deutsch}}, \bibnamefont{and}
  \bibinfo{author}{\bibfnamefont{M.}~\bibnamefont{Lewenstein}},
  \bibinfo{journal}{Reports on Progress in Physics}
  \textbf{\bibinfo{volume}{75}}, \bibinfo{pages}{082401}
  (\bibinfo{year}{2012}),
  \urlprefix\url{https://doi.org/10.1088%2F0034-4885%2F75%2F8%2F082401}.

\bibitem[{\citenamefont{Georgescu et~al.}(2014)\citenamefont{Georgescu, Ashhab,
  and Nori}}]{Georgescu2014}
\bibinfo{author}{\bibfnamefont{I.~M.} \bibnamefont{Georgescu}},
  \bibinfo{author}{\bibfnamefont{S.}~\bibnamefont{Ashhab}}, \bibnamefont{and}
  \bibinfo{author}{\bibfnamefont{F.}~\bibnamefont{Nori}},
  \bibinfo{journal}{Rev. Mod. Phys.} \textbf{\bibinfo{volume}{86}},
  \bibinfo{pages}{153} (\bibinfo{year}{2014}),
  \urlprefix\url{https://link.aps.org/doi/10.1103/RevModPhys.86.153}.

\bibitem[{\citenamefont{Amico et~al.}(2008)\citenamefont{Amico, Fazio,
  Osterloh, and Vedral}}]{Amico2008}
\bibinfo{author}{\bibfnamefont{L.}~\bibnamefont{Amico}},
  \bibinfo{author}{\bibfnamefont{R.}~\bibnamefont{Fazio}},
  \bibinfo{author}{\bibfnamefont{A.}~\bibnamefont{Osterloh}}, \bibnamefont{and}
  \bibinfo{author}{\bibfnamefont{V.}~\bibnamefont{Vedral}},
  \bibinfo{journal}{Rev. Mod. Phys.} \textbf{\bibinfo{volume}{80}},
  \bibinfo{pages}{517} (\bibinfo{year}{2008}),
  \urlprefix\url{https://link.aps.org/doi/10.1103/RevModPhys.80.517}.

\bibitem[{\citenamefont{Laflorencie}(2016)}]{Laflorencie2016}
\bibinfo{author}{\bibfnamefont{N.}~\bibnamefont{Laflorencie}},
  \bibinfo{journal}{Physics Reports} \textbf{\bibinfo{volume}{646}},
  \bibinfo{pages}{1 } (\bibinfo{year}{2016}), ISSN \bibinfo{issn}{0370-1573},
  \bibinfo{note}{quantum entanglement in condensed matter systems},
  \urlprefix\url{http://www.sciencedirect.com/science/article/pii/S0370157316301582}.

\bibitem[{\citenamefont{Holzhey et~al.}(1994)\citenamefont{Holzhey, Larsen, and
  Wilczek}}]{Holzhey1994}
\bibinfo{author}{\bibfnamefont{C.}~\bibnamefont{Holzhey}},
  \bibinfo{author}{\bibfnamefont{F.}~\bibnamefont{Larsen}}, \bibnamefont{and}
  \bibinfo{author}{\bibfnamefont{F.}~\bibnamefont{Wilczek}},
  \bibinfo{journal}{Nuclear Physics B} \textbf{\bibinfo{volume}{424}},
  \bibinfo{pages}{443 } (\bibinfo{year}{1994}), ISSN \bibinfo{issn}{0550-3213},
  \urlprefix\url{http://www.sciencedirect.com/science/article/pii/0550321394904022}.

\bibitem[{\citenamefont{Calabrese and Cardy}(2004)}]{calabrese2004entanglement}
\bibinfo{author}{\bibfnamefont{P.}~\bibnamefont{Calabrese}} \bibnamefont{and}
  \bibinfo{author}{\bibfnamefont{J.}~\bibnamefont{Cardy}},
  \bibinfo{journal}{Journal of Statistical Mechanics: Theory and Experiment}
  \textbf{\bibinfo{volume}{2004}}, \bibinfo{pages}{P06002}
  (\bibinfo{year}{2004}).

\bibitem[{\citenamefont{Eisert et~al.}(2010)\citenamefont{Eisert, Cramer, and
  Plenio}}]{Eisert2010}
\bibinfo{author}{\bibfnamefont{J.}~\bibnamefont{Eisert}},
  \bibinfo{author}{\bibfnamefont{M.}~\bibnamefont{Cramer}}, \bibnamefont{and}
  \bibinfo{author}{\bibfnamefont{M.~B.} \bibnamefont{Plenio}},
  \bibinfo{journal}{Rev. Mod. Phys.} \textbf{\bibinfo{volume}{82}},
  \bibinfo{pages}{277} (\bibinfo{year}{2010}),
  \urlprefix\url{https://link.aps.org/doi/10.1103/RevModPhys.82.277}.

\bibitem[{\citenamefont{Calabrese and Cardy}(2005)}]{Calabrese2005}
\bibinfo{author}{\bibfnamefont{P.}~\bibnamefont{Calabrese}} \bibnamefont{and}
  \bibinfo{author}{\bibfnamefont{J.}~\bibnamefont{Cardy}},
  \bibinfo{journal}{Journal of Statistical Mechanics: Theory and Experiment}
  \textbf{\bibinfo{volume}{2005}}, \bibinfo{pages}{P04010}
  (\bibinfo{year}{2005}),
  \urlprefix\url{https://doi.org/10.1088%2F1742-5468%2F2005%2F04%2Fp04010}.

\bibitem[{\citenamefont{Fagotti and Calabrese}(2008)}]{Fagotti2008}
\bibinfo{author}{\bibfnamefont{M.}~\bibnamefont{Fagotti}} \bibnamefont{and}
  \bibinfo{author}{\bibfnamefont{P.}~\bibnamefont{Calabrese}},
  \bibinfo{journal}{Phys. Rev. A} \textbf{\bibinfo{volume}{78}},
  \bibinfo{pages}{010306} (\bibinfo{year}{2008}),
  \urlprefix\url{https://link.aps.org/doi/10.1103/PhysRevA.78.010306}.

\bibitem[{\citenamefont{Nezhadhaghighi and
  Rajabpour}(2014)}]{Nezhadhaghighi2014}
\bibinfo{author}{\bibfnamefont{M.~G.} \bibnamefont{Nezhadhaghighi}}
  \bibnamefont{and} \bibinfo{author}{\bibfnamefont{M.~A.}
  \bibnamefont{Rajabpour}}, \bibinfo{journal}{Phys. Rev. B}
  \textbf{\bibinfo{volume}{90}}, \bibinfo{pages}{205438}
  (\bibinfo{year}{2014}),
  \urlprefix\url{https://link.aps.org/doi/10.1103/PhysRevB.90.205438}.

\bibitem[{\citenamefont{Bucciantini et~al.}(2014)\citenamefont{Bucciantini,
  Kormos, and Calabrese}}]{Bucciantini_2014}
\bibinfo{author}{\bibfnamefont{L.}~\bibnamefont{Bucciantini}},
  \bibinfo{author}{\bibfnamefont{M.}~\bibnamefont{Kormos}}, \bibnamefont{and}
  \bibinfo{author}{\bibfnamefont{P.}~\bibnamefont{Calabrese}},
  \bibinfo{journal}{Journal of Physics A: Mathematical and Theoretical}
  \textbf{\bibinfo{volume}{47}}, \bibinfo{pages}{175002}
  (\bibinfo{year}{2014}),
  \urlprefix\url{https://doi.org/10.1088%2F1751-8113%2F47%2F17%2F175002}.

\bibitem[{\citenamefont{Cardy and Tonni}(2016)}]{Cardy_2016}
\bibinfo{author}{\bibfnamefont{J.}~\bibnamefont{Cardy}} \bibnamefont{and}
  \bibinfo{author}{\bibfnamefont{E.}~\bibnamefont{Tonni}},
  \bibinfo{journal}{Journal of Statistical Mechanics: Theory and Experiment}
  \textbf{\bibinfo{volume}{2016}}, \bibinfo{pages}{123103}
  (\bibinfo{year}{2016}),
  \urlprefix\url{https://doi.org/10.1088%2F1742-5468%2F2016%2F12%2F123103}.

\bibitem[{\citenamefont{Cotler et~al.}(2016)\citenamefont{Cotler, Hertzberg,
  Mezei, and Mueller}}]{cotler2016entanglement}
\bibinfo{author}{\bibfnamefont{J.~S.} \bibnamefont{Cotler}},
  \bibinfo{author}{\bibfnamefont{M.~P.} \bibnamefont{Hertzberg}},
  \bibinfo{author}{\bibfnamefont{M.}~\bibnamefont{Mezei}}, \bibnamefont{and}
  \bibinfo{author}{\bibfnamefont{M.~T.} \bibnamefont{Mueller}},
  \bibinfo{journal}{Journal of High Energy Physics}
  \textbf{\bibinfo{volume}{2016}}, \bibinfo{pages}{166} (\bibinfo{year}{2016}).

\bibitem[{\citenamefont{Wen et~al.}(2018)\citenamefont{Wen, Ryu, and
  Ludwig}}]{Wen_2018}
\bibinfo{author}{\bibfnamefont{X.}~\bibnamefont{Wen}},
  \bibinfo{author}{\bibfnamefont{S.}~\bibnamefont{Ryu}}, \bibnamefont{and}
  \bibinfo{author}{\bibfnamefont{A.~W.~W.} \bibnamefont{Ludwig}},
  \bibinfo{journal}{Journal of Statistical Mechanics: Theory and Experiment}
  \textbf{\bibinfo{volume}{2018}}, \bibinfo{pages}{113103}
  (\bibinfo{year}{2018}),
  \urlprefix\url{https://doi.org/10.1088%2F1742-5468%2Faae84e}.

\bibitem[{\citenamefont{Najafi et~al.}(2018)\citenamefont{Najafi, Rajabpour,
  and Viti}}]{Najafi2018}
\bibinfo{author}{\bibfnamefont{K.}~\bibnamefont{Najafi}},
  \bibinfo{author}{\bibfnamefont{M.~A.} \bibnamefont{Rajabpour}},
  \bibnamefont{and} \bibinfo{author}{\bibfnamefont{J.}~\bibnamefont{Viti}},
  \bibinfo{journal}{Phys. Rev. B} \textbf{\bibinfo{volume}{97}},
  \bibinfo{pages}{205103} (\bibinfo{year}{2018}),
  \urlprefix\url{https://link.aps.org/doi/10.1103/PhysRevB.97.205103}.

\bibitem[{\citenamefont{Kim and Huse}(2013)}]{Kim2013}
\bibinfo{author}{\bibfnamefont{H.}~\bibnamefont{Kim}} \bibnamefont{and}
  \bibinfo{author}{\bibfnamefont{D.~A.} \bibnamefont{Huse}},
  \bibinfo{journal}{Phys. Rev. Lett.} \textbf{\bibinfo{volume}{111}},
  \bibinfo{pages}{127205} (\bibinfo{year}{2013}),
  \urlprefix\url{https://link.aps.org/doi/10.1103/PhysRevLett.111.127205}.

\bibitem[{\citenamefont{Buyskikh et~al.}(2016)\citenamefont{Buyskikh, Fagotti,
  Schachenmayer, Essler, and Daley}}]{Buyskikh2016}
\bibinfo{author}{\bibfnamefont{A.~S.} \bibnamefont{Buyskikh}},
  \bibinfo{author}{\bibfnamefont{M.}~\bibnamefont{Fagotti}},
  \bibinfo{author}{\bibfnamefont{J.}~\bibnamefont{Schachenmayer}},
  \bibinfo{author}{\bibfnamefont{F.}~\bibnamefont{Essler}}, \bibnamefont{and}
  \bibinfo{author}{\bibfnamefont{A.~J.} \bibnamefont{Daley}},
  \bibinfo{journal}{Phys. Rev. A} \textbf{\bibinfo{volume}{93}},
  \bibinfo{pages}{053620} (\bibinfo{year}{2016}),
  \urlprefix\url{https://link.aps.org/doi/10.1103/PhysRevA.93.053620}.

\bibitem[{\citenamefont{Alba and Calabrese}(2017)}]{Alba2017}
\bibinfo{author}{\bibfnamefont{V.}~\bibnamefont{Alba}} \bibnamefont{and}
  \bibinfo{author}{\bibfnamefont{P.}~\bibnamefont{Calabrese}},
  \bibinfo{journal}{Proceedings of the National Academy of Sciences}
  \textbf{\bibinfo{volume}{114}}, \bibinfo{pages}{7947} (\bibinfo{year}{2017}),
  ISSN \bibinfo{issn}{0027-8424},
  \eprint{https://www.pnas.org/content/114/30/7947.full.pdf},
  \urlprefix\url{https://www.pnas.org/content/114/30/7947}.

\bibitem[{\citenamefont{Alba and Calabrese}(2018)}]{Alba2018}
\bibinfo{author}{\bibfnamefont{V.}~\bibnamefont{Alba}} \bibnamefont{and}
  \bibinfo{author}{\bibfnamefont{P.}~\bibnamefont{Calabrese}},
  \bibinfo{journal}{SciPost Phys.} \textbf{\bibinfo{volume}{4}},
  \bibinfo{pages}{17} (\bibinfo{year}{2018}),
  \urlprefix\url{https://scipost.org/10.21468/SciPostPhys.4.3.017}.

\bibitem[{\citenamefont{Alba}(2018)}]{Alba2018b}
\bibinfo{author}{\bibfnamefont{V.}~\bibnamefont{Alba}}, \bibinfo{journal}{Phys.
  Rev. B} \textbf{\bibinfo{volume}{97}}, \bibinfo{pages}{245135}
  (\bibinfo{year}{2018}),
  \urlprefix\url{https://link.aps.org/doi/10.1103/PhysRevB.97.245135}.

\bibitem[{\citenamefont{Bertini
  et~al.}(2018{\natexlab{a}})\citenamefont{Bertini, Tartaglia, and
  Calabrese}}]{Bertini_2018}
\bibinfo{author}{\bibfnamefont{B.}~\bibnamefont{Bertini}},
  \bibinfo{author}{\bibfnamefont{E.}~\bibnamefont{Tartaglia}},
  \bibnamefont{and}
  \bibinfo{author}{\bibfnamefont{P.}~\bibnamefont{Calabrese}},
  \bibinfo{journal}{Journal of Statistical Mechanics: Theory and Experiment}
  \textbf{\bibinfo{volume}{2018}}, \bibinfo{pages}{063104}
  (\bibinfo{year}{2018}{\natexlab{a}}),
  \urlprefix\url{https://doi.org/10.1088%2F1742-5468%2Faac73f}.

\bibitem[{\citenamefont{Viti et~al.}(2016)\citenamefont{Viti, St{\'{e}}phan,
  Dubail, and Haque}}]{Viti_2016}
\bibinfo{author}{\bibfnamefont{J.}~\bibnamefont{Viti}},
  \bibinfo{author}{\bibfnamefont{J.-M.} \bibnamefont{St{\'{e}}phan}},
  \bibinfo{author}{\bibfnamefont{J.}~\bibnamefont{Dubail}}, \bibnamefont{and}
  \bibinfo{author}{\bibfnamefont{M.}~\bibnamefont{Haque}},
  \bibinfo{journal}{{EPL} (Europhysics Letters)}
  \textbf{\bibinfo{volume}{115}}, \bibinfo{pages}{40011}
  (\bibinfo{year}{2016}),
  \urlprefix\url{https://doi.org/10.1209%2F0295-5075%2F115%2F40011}.

\bibitem[{\citenamefont{Dubail et~al.}(2017)\citenamefont{Dubail,
  St{\'{e}}phan, Viti, and Calabrese}}]{Dubail2017}
\bibinfo{author}{\bibfnamefont{J.}~\bibnamefont{Dubail}},
  \bibinfo{author}{\bibfnamefont{J.-M.} \bibnamefont{St{\'{e}}phan}},
  \bibinfo{author}{\bibfnamefont{J.}~\bibnamefont{Viti}}, \bibnamefont{and}
  \bibinfo{author}{\bibfnamefont{P.}~\bibnamefont{Calabrese}},
  \bibinfo{journal}{SciPost Phys.} \textbf{\bibinfo{volume}{2}},
  \bibinfo{pages}{002} (\bibinfo{year}{2017}),
  \urlprefix\url{https://scipost.org/10.21468/SciPostPhys.2.1.002}.

\bibitem[{\citenamefont{Bertini
  et~al.}(2018{\natexlab{b}})\citenamefont{Bertini, Fagotti, Piroli, and
  Calabrese}}]{Bertini2018b}
\bibinfo{author}{\bibfnamefont{B.}~\bibnamefont{Bertini}},
  \bibinfo{author}{\bibfnamefont{M.}~\bibnamefont{Fagotti}},
  \bibinfo{author}{\bibfnamefont{L.}~\bibnamefont{Piroli}}, \bibnamefont{and}
  \bibinfo{author}{\bibfnamefont{P.}~\bibnamefont{Calabrese}},
  \bibinfo{journal}{Journal of Physics A: Mathematical and Theoretical}
  \textbf{\bibinfo{volume}{51}}, \bibinfo{pages}{39LT01}
  (\bibinfo{year}{2018}{\natexlab{b}}),
  \urlprefix\url{https://doi.org/10.1088%2F1751-8121%2Faad82e}.

\bibitem[{\citenamefont{Alba et~al.}(2019)\citenamefont{Alba, Bertini, and
  Fagotti}}]{Alba2019}
\bibinfo{author}{\bibfnamefont{V.}~\bibnamefont{Alba}},
  \bibinfo{author}{\bibfnamefont{B.}~\bibnamefont{Bertini}}, \bibnamefont{and}
  \bibinfo{author}{\bibfnamefont{M.}~\bibnamefont{Fagotti}},
  \bibinfo{journal}{SciPost Phys.} \textbf{\bibinfo{volume}{7}},
  \bibinfo{pages}{5} (\bibinfo{year}{2019}),
  \urlprefix\url{https://scipost.org/10.21468/SciPostPhys.7.1.005}.

\bibitem[{\citenamefont{Bardarson et~al.}(2012)\citenamefont{Bardarson,
  Pollmann, and Moore}}]{Bardarson2012}
\bibinfo{author}{\bibfnamefont{J.~H.} \bibnamefont{Bardarson}},
  \bibinfo{author}{\bibfnamefont{F.}~\bibnamefont{Pollmann}}, \bibnamefont{and}
  \bibinfo{author}{\bibfnamefont{J.~E.} \bibnamefont{Moore}},
  \bibinfo{journal}{Phys. Rev. Lett.} \textbf{\bibinfo{volume}{109}},
  \bibinfo{pages}{017202} (\bibinfo{year}{2012}),
  \urlprefix\url{https://link.aps.org/doi/10.1103/PhysRevLett.109.017202}.

\bibitem[{\citenamefont{Serbyn et~al.}(2013)\citenamefont{Serbyn, Papi{\'c},
  and Abanin}}]{serbyn2013universal}
\bibinfo{author}{\bibfnamefont{M.}~\bibnamefont{Serbyn}},
  \bibinfo{author}{\bibfnamefont{Z.}~\bibnamefont{Papi{\'c}}},
  \bibnamefont{and} \bibinfo{author}{\bibfnamefont{D.~A.}
  \bibnamefont{Abanin}}, \bibinfo{journal}{Physical review letters}
  \textbf{\bibinfo{volume}{110}}, \bibinfo{pages}{260601}
  (\bibinfo{year}{2013}).

\bibitem[{\citenamefont{Huse et~al.}(2014)\citenamefont{Huse, Nandkishore, and
  Oganesyan}}]{huse2014phenomenology}
\bibinfo{author}{\bibfnamefont{D.~A.} \bibnamefont{Huse}},
  \bibinfo{author}{\bibfnamefont{R.}~\bibnamefont{Nandkishore}},
  \bibnamefont{and}
  \bibinfo{author}{\bibfnamefont{V.}~\bibnamefont{Oganesyan}},
  \bibinfo{journal}{Physical Review B} \textbf{\bibinfo{volume}{90}},
  \bibinfo{pages}{174202} (\bibinfo{year}{2014}).

\bibitem[{\citenamefont{Schreiber et~al.}(2015)\citenamefont{Schreiber,
  Hodgman, Bordia, L{\"u}schen, Fischer, Vosk, Altman, Schneider, and
  Bloch}}]{Schreiber2015}
\bibinfo{author}{\bibfnamefont{M.}~\bibnamefont{Schreiber}},
  \bibinfo{author}{\bibfnamefont{S.~S.} \bibnamefont{Hodgman}},
  \bibinfo{author}{\bibfnamefont{P.}~\bibnamefont{Bordia}},
  \bibinfo{author}{\bibfnamefont{H.~P.} \bibnamefont{L{\"u}schen}},
  \bibinfo{author}{\bibfnamefont{M.~H.} \bibnamefont{Fischer}},
  \bibinfo{author}{\bibfnamefont{R.}~\bibnamefont{Vosk}},
  \bibinfo{author}{\bibfnamefont{E.}~\bibnamefont{Altman}},
  \bibinfo{author}{\bibfnamefont{U.}~\bibnamefont{Schneider}},
  \bibnamefont{and} \bibinfo{author}{\bibfnamefont{I.}~\bibnamefont{Bloch}},
  \bibinfo{journal}{Science} \textbf{\bibinfo{volume}{349}},
  \bibinfo{pages}{842} (\bibinfo{year}{2015}), ISSN \bibinfo{issn}{0036-8075},
  \eprint{https://science.sciencemag.org/content/349/6250/842.full.pdf},
  \urlprefix\url{https://science.sciencemag.org/content/349/6250/842}.

\bibitem[{\citenamefont{Abanin et~al.}(2019)\citenamefont{Abanin, Altman,
  Bloch, and Serbyn}}]{Abanin2019}
\bibinfo{author}{\bibfnamefont{D.~A.} \bibnamefont{Abanin}},
  \bibinfo{author}{\bibfnamefont{E.}~\bibnamefont{Altman}},
  \bibinfo{author}{\bibfnamefont{I.}~\bibnamefont{Bloch}}, \bibnamefont{and}
  \bibinfo{author}{\bibfnamefont{M.}~\bibnamefont{Serbyn}},
  \bibinfo{journal}{Rev. Mod. Phys.} \textbf{\bibinfo{volume}{91}},
  \bibinfo{pages}{021001} (\bibinfo{year}{2019}),
  \urlprefix\url{https://link.aps.org/doi/10.1103/RevModPhys.91.021001}.

\bibitem[{\citenamefont{Birrell and Davies}(1982)}]{birrell_davies_1982}
\bibinfo{author}{\bibfnamefont{N.~D.} \bibnamefont{Birrell}} \bibnamefont{and}
  \bibinfo{author}{\bibfnamefont{P.~C.~W.} \bibnamefont{Davies}},
  \emph{\bibinfo{title}{Quantum Fields in Curved Space}}, Cambridge Monographs
  on Mathematical Physics (\bibinfo{publisher}{Cambridge University Press},
  \bibinfo{year}{1982}).

\bibitem[{\citenamefont{Takagi}(1986)}]{takagi1986vacuum}
\bibinfo{author}{\bibfnamefont{S.}~\bibnamefont{Takagi}},
  \bibinfo{journal}{Progress of Theoretical Physics Supplement}
  \textbf{\bibinfo{volume}{88}}, \bibinfo{pages}{1} (\bibinfo{year}{1986}).

\bibitem[{\citenamefont{Wald}(2010)}]{wald2010general}
\bibinfo{author}{\bibfnamefont{R.~M.} \bibnamefont{Wald}},
  \emph{\bibinfo{title}{General relativity}} (\bibinfo{publisher}{University of
  Chicago press}, \bibinfo{year}{2010}).

\bibitem[{\citenamefont{Nahum et~al.}(2017)\citenamefont{Nahum, Ruhman, Vijay,
  and Haah}}]{nahum2017quantum}
\bibinfo{author}{\bibfnamefont{A.}~\bibnamefont{Nahum}},
  \bibinfo{author}{\bibfnamefont{J.}~\bibnamefont{Ruhman}},
  \bibinfo{author}{\bibfnamefont{S.}~\bibnamefont{Vijay}}, \bibnamefont{and}
  \bibinfo{author}{\bibfnamefont{J.}~\bibnamefont{Haah}},
  \bibinfo{journal}{Physical Review X} \textbf{\bibinfo{volume}{7}},
  \bibinfo{pages}{031016} (\bibinfo{year}{2017}).

\bibitem[{\citenamefont{Jonay et~al.}(2018)\citenamefont{Jonay, Huse, and
  Nahum}}]{jonay2018coarsegrained}
\bibinfo{author}{\bibfnamefont{C.}~\bibnamefont{Jonay}},
  \bibinfo{author}{\bibfnamefont{D.~A.} \bibnamefont{Huse}}, \bibnamefont{and}
  \bibinfo{author}{\bibfnamefont{A.}~\bibnamefont{Nahum}},
  \emph{\bibinfo{title}{Coarse-grained dynamics of operator and state
  entanglement}} (\bibinfo{year}{2018}), \eprint{1803.00089}.

\bibitem[{\citenamefont{Nahum et~al.}(2018{\natexlab{a}})\citenamefont{Nahum,
  Vijay, and Haah}}]{Nahum2018}
\bibinfo{author}{\bibfnamefont{A.}~\bibnamefont{Nahum}},
  \bibinfo{author}{\bibfnamefont{S.}~\bibnamefont{Vijay}}, \bibnamefont{and}
  \bibinfo{author}{\bibfnamefont{J.}~\bibnamefont{Haah}},
  \bibinfo{journal}{Phys. Rev. X} \textbf{\bibinfo{volume}{8}},
  \bibinfo{pages}{021014} (\bibinfo{year}{2018}{\natexlab{a}}),
  \urlprefix\url{https://link.aps.org/doi/10.1103/PhysRevX.8.021014}.

\bibitem[{\citenamefont{Nahum et~al.}(2018{\natexlab{b}})\citenamefont{Nahum,
  Ruhman, and Huse}}]{Nahum2018b}
\bibinfo{author}{\bibfnamefont{A.}~\bibnamefont{Nahum}},
  \bibinfo{author}{\bibfnamefont{J.}~\bibnamefont{Ruhman}}, \bibnamefont{and}
  \bibinfo{author}{\bibfnamefont{D.~A.} \bibnamefont{Huse}},
  \bibinfo{journal}{Phys. Rev. B} \textbf{\bibinfo{volume}{98}},
  \bibinfo{pages}{035118} (\bibinfo{year}{2018}{\natexlab{b}}),
  \urlprefix\url{https://link.aps.org/doi/10.1103/PhysRevB.98.035118}.

\bibitem[{\citenamefont{Plenio and Virmani}(2014)}]{plenio2014introduction}
\bibinfo{author}{\bibfnamefont{M.~B.} \bibnamefont{Plenio}} \bibnamefont{and}
  \bibinfo{author}{\bibfnamefont{S.~S.} \bibnamefont{Virmani}}, in
  \emph{\bibinfo{booktitle}{Quantum Information and Coherence}}
  (\bibinfo{publisher}{Springer}, \bibinfo{year}{2014}), pp.
  \bibinfo{pages}{173--209}.

\bibitem[{\citenamefont{Maity et~al.}(2020)\citenamefont{Maity, Bandyopadhyay,
  Bhattacharjee, and Dutta}}]{Maity2020}
\bibinfo{author}{\bibfnamefont{S.}~\bibnamefont{Maity}},
  \bibinfo{author}{\bibfnamefont{S.}~\bibnamefont{Bandyopadhyay}},
  \bibinfo{author}{\bibfnamefont{S.}~\bibnamefont{Bhattacharjee}},
  \bibnamefont{and} \bibinfo{author}{\bibfnamefont{A.}~\bibnamefont{Dutta}},
  \bibinfo{journal}{Phys. Rev. B} \textbf{\bibinfo{volume}{101}},
  \bibinfo{pages}{180301} (\bibinfo{year}{2020}),
  \urlprefix\url{https://link.aps.org/doi/10.1103/PhysRevB.101.180301}.

\bibitem[{\citenamefont{Lieb and Robinson}(1972)}]{lieb1972finite}
\bibinfo{author}{\bibfnamefont{E.~H.} \bibnamefont{Lieb}} \bibnamefont{and}
  \bibinfo{author}{\bibfnamefont{D.~W.} \bibnamefont{Robinson}}, in
  \emph{\bibinfo{booktitle}{Statistical mechanics}}
  (\bibinfo{publisher}{Springer}, \bibinfo{year}{1972}), pp.
  \bibinfo{pages}{425--431}.

\bibitem[{\citenamefont{Wen and Wu}(2018)}]{wen2018}
\bibinfo{author}{\bibfnamefont{X.}~\bibnamefont{Wen}} \bibnamefont{and}
  \bibinfo{author}{\bibfnamefont{J.-Q.} \bibnamefont{Wu}},
  \emph{\bibinfo{title}{Floquet conformal field theory}}
  (\bibinfo{year}{2018}), \eprint{1805.00031}.

\bibitem[{\citenamefont{Lapierre et~al.}(2020)\citenamefont{Lapierre, Choo,
  Tauber, Tiwari, Neupert, and Chitra}}]{Lapierre2020}
\bibinfo{author}{\bibfnamefont{B.}~\bibnamefont{Lapierre}},
  \bibinfo{author}{\bibfnamefont{K.}~\bibnamefont{Choo}},
  \bibinfo{author}{\bibfnamefont{C.}~\bibnamefont{Tauber}},
  \bibinfo{author}{\bibfnamefont{A.}~\bibnamefont{Tiwari}},
  \bibinfo{author}{\bibfnamefont{T.}~\bibnamefont{Neupert}}, \bibnamefont{and}
  \bibinfo{author}{\bibfnamefont{R.}~\bibnamefont{Chitra}},
  \bibinfo{journal}{Phys. Rev. Research} \textbf{\bibinfo{volume}{2}},
  \bibinfo{pages}{023085} (\bibinfo{year}{2020}),
  \urlprefix\url{https://link.aps.org/doi/10.1103/PhysRevResearch.2.023085}.

\bibitem[{\citenamefont{Schwarzschild}(1916)}]{schwarzschild1916gravitational}
\bibinfo{author}{\bibfnamefont{K.}~\bibnamefont{Schwarzschild}},
  \bibinfo{journal}{Abh. Konigl. Preuss. Akad. Wissenschaften Jahre 1906, 92,
  Berlin, 1907} \textbf{\bibinfo{volume}{1916}} (\bibinfo{year}{1916}).

\bibitem[{\citenamefont{Unruh}(1976)}]{Unruh1976}
\bibinfo{author}{\bibfnamefont{W.~G.} \bibnamefont{Unruh}},
  \bibinfo{journal}{Phys. Rev. D} \textbf{\bibinfo{volume}{14}},
  \bibinfo{pages}{870} (\bibinfo{year}{1976}),
  \urlprefix\url{https://link.aps.org/doi/10.1103/PhysRevD.14.870}.

\bibitem[{\citenamefont{Boada et~al.}(2011)\citenamefont{Boada, Celi, Latorre,
  and Lewenstein}}]{Boada_2011}
\bibinfo{author}{\bibfnamefont{O.}~\bibnamefont{Boada}},
  \bibinfo{author}{\bibfnamefont{A.}~\bibnamefont{Celi}},
  \bibinfo{author}{\bibfnamefont{J.~I.} \bibnamefont{Latorre}},
  \bibnamefont{and}
  \bibinfo{author}{\bibfnamefont{M.}~\bibnamefont{Lewenstein}},
  \bibinfo{journal}{New Journal of Physics} \textbf{\bibinfo{volume}{13}},
  \bibinfo{pages}{035002} (\bibinfo{year}{2011}),
  \urlprefix\url{https://doi.org/10.1088%2F1367-2630%2F13%2F3%2F035002}.

\bibitem[{\citenamefont{Celi}(2017)}]{celi2017different}
\bibinfo{author}{\bibfnamefont{A.}~\bibnamefont{Celi}}, \bibinfo{journal}{The
  European Physical Journal Special Topics} \textbf{\bibinfo{volume}{226}},
  \bibinfo{pages}{2729} (\bibinfo{year}{2017}).

\bibitem[{\citenamefont{Rodr{\'\i}guez-Laguna
  et~al.}(2017)\citenamefont{Rodr{\'\i}guez-Laguna, Tarruell, Lewenstein, and
  Celi}}]{rodriguez2017synthetic}
\bibinfo{author}{\bibfnamefont{J.}~\bibnamefont{Rodr{\'\i}guez-Laguna}},
  \bibinfo{author}{\bibfnamefont{L.}~\bibnamefont{Tarruell}},
  \bibinfo{author}{\bibfnamefont{M.}~\bibnamefont{Lewenstein}},
  \bibnamefont{and} \bibinfo{author}{\bibfnamefont{A.}~\bibnamefont{Celi}},
  \bibinfo{journal}{Physical Review A} \textbf{\bibinfo{volume}{95}},
  \bibinfo{pages}{013627} (\bibinfo{year}{2017}).

\bibitem[{\citenamefont{Kosior et~al.}(2018)\citenamefont{Kosior, Lewenstein,
  and Celi}}]{kosior2018}
\bibinfo{author}{\bibfnamefont{A.}~\bibnamefont{Kosior}},
  \bibinfo{author}{\bibfnamefont{M.}~\bibnamefont{Lewenstein}},
  \bibnamefont{and} \bibinfo{author}{\bibfnamefont{A.}~\bibnamefont{Celi}},
  \bibinfo{journal}{SciPost Phys.} \textbf{\bibinfo{volume}{5}},
  \bibinfo{pages}{61} (\bibinfo{year}{2018}),
  \urlprefix\url{https://scipost.org/10.21468/SciPostPhys.5.6.061}.

\bibitem[{\citenamefont{Louko}(2018)}]{louko2018thermality}
\bibinfo{author}{\bibfnamefont{J.}~\bibnamefont{Louko}},
  \bibinfo{journal}{Classical and Quantum Gravity}
  \textbf{\bibinfo{volume}{35}}, \bibinfo{pages}{205006}
  (\bibinfo{year}{2018}).

\bibitem[{\citenamefont{Yang et~al.}(2020)\citenamefont{Yang, Liu, Zhu, Luo,
  and Cai}}]{Yang2020}
\bibinfo{author}{\bibfnamefont{R.-Q.} \bibnamefont{Yang}},
  \bibinfo{author}{\bibfnamefont{H.}~\bibnamefont{Liu}},
  \bibinfo{author}{\bibfnamefont{S.}~\bibnamefont{Zhu}},
  \bibinfo{author}{\bibfnamefont{L.}~\bibnamefont{Luo}}, \bibnamefont{and}
  \bibinfo{author}{\bibfnamefont{R.-G.} \bibnamefont{Cai}},
  \bibinfo{journal}{Phys. Rev. Research} \textbf{\bibinfo{volume}{2}},
  \bibinfo{pages}{023107} (\bibinfo{year}{2020}),
  \urlprefix\url{https://link.aps.org/doi/10.1103/PhysRevResearch.2.023107}.

\bibitem[{\citenamefont{Dalmonte et~al.}(2018)\citenamefont{Dalmonte,
  Vermersch, and Zoller}}]{dalmonte2018quantum}
\bibinfo{author}{\bibfnamefont{M.}~\bibnamefont{Dalmonte}},
  \bibinfo{author}{\bibfnamefont{B.}~\bibnamefont{Vermersch}},
  \bibnamefont{and} \bibinfo{author}{\bibfnamefont{P.}~\bibnamefont{Zoller}},
  \bibinfo{journal}{Nature Physics} \textbf{\bibinfo{volume}{14}},
  \bibinfo{pages}{827} (\bibinfo{year}{2018}).

\bibitem[{\citenamefont{Giudici et~al.}(2018)\citenamefont{Giudici,
  Mendes-Santos, Calabrese, and Dalmonte}}]{Giudici2018}
\bibinfo{author}{\bibfnamefont{G.}~\bibnamefont{Giudici}},
  \bibinfo{author}{\bibfnamefont{T.}~\bibnamefont{Mendes-Santos}},
  \bibinfo{author}{\bibfnamefont{P.}~\bibnamefont{Calabrese}},
  \bibnamefont{and} \bibinfo{author}{\bibfnamefont{M.}~\bibnamefont{Dalmonte}},
  \bibinfo{journal}{Phys. Rev. B} \textbf{\bibinfo{volume}{98}},
  \bibinfo{pages}{134403} (\bibinfo{year}{2018}),
  \urlprefix\url{https://link.aps.org/doi/10.1103/PhysRevB.98.134403}.

\bibitem[{\citenamefont{Eisler and Peschel}(2018)}]{Eisler_2018}
\bibinfo{author}{\bibfnamefont{V.}~\bibnamefont{Eisler}} \bibnamefont{and}
  \bibinfo{author}{\bibfnamefont{I.}~\bibnamefont{Peschel}},
  \bibinfo{journal}{Journal of Statistical Mechanics: Theory and Experiment}
  \textbf{\bibinfo{volume}{2018}}, \bibinfo{pages}{104001}
  (\bibinfo{year}{2018}),
  \urlprefix\url{https://doi.org/10.1088%2F1742-5468%2Faace2b}.

\bibitem[{\citenamefont{Parisen~Toldin and Assaad}(2018)}]{Parisen2018}
\bibinfo{author}{\bibfnamefont{F.}~\bibnamefont{Parisen~Toldin}}
  \bibnamefont{and} \bibinfo{author}{\bibfnamefont{F.~F.}
  \bibnamefont{Assaad}}, \bibinfo{journal}{Phys. Rev. Lett.}
  \textbf{\bibinfo{volume}{121}}, \bibinfo{pages}{200602}
  (\bibinfo{year}{2018}),
  \urlprefix\url{https://link.aps.org/doi/10.1103/PhysRevLett.121.200602}.

\bibitem[{\citenamefont{Zhu et~al.}(2019)\citenamefont{Zhu, Huang, and
  He}}]{Zhu2019}
\bibinfo{author}{\bibfnamefont{W.}~\bibnamefont{Zhu}},
  \bibinfo{author}{\bibfnamefont{Z.}~\bibnamefont{Huang}}, \bibnamefont{and}
  \bibinfo{author}{\bibfnamefont{Y.-C.} \bibnamefont{He}},
  \bibinfo{journal}{Phys. Rev. B} \textbf{\bibinfo{volume}{99}},
  \bibinfo{pages}{235109} (\bibinfo{year}{2019}),
  \urlprefix\url{https://link.aps.org/doi/10.1103/PhysRevB.99.235109}.

\bibitem[{\citenamefont{Turkeshi et~al.}(2019)\citenamefont{Turkeshi,
  Mendes-Santos, Giudici, and Dalmonte}}]{Turkeshi2019}
\bibinfo{author}{\bibfnamefont{X.}~\bibnamefont{Turkeshi}},
  \bibinfo{author}{\bibfnamefont{T.}~\bibnamefont{Mendes-Santos}},
  \bibinfo{author}{\bibfnamefont{G.}~\bibnamefont{Giudici}}, \bibnamefont{and}
  \bibinfo{author}{\bibfnamefont{M.}~\bibnamefont{Dalmonte}},
  \bibinfo{journal}{Phys. Rev. Lett.} \textbf{\bibinfo{volume}{122}},
  \bibinfo{pages}{150606} (\bibinfo{year}{2019}),
  \urlprefix\url{https://link.aps.org/doi/10.1103/PhysRevLett.122.150606}.

\bibitem[{\citenamefont{Mendes-Santos et~al.}(2019)\citenamefont{Mendes-Santos,
  Giudici, Dalmonte, and Rajabpour}}]{MendesSantos2019}
\bibinfo{author}{\bibfnamefont{T.}~\bibnamefont{Mendes-Santos}},
  \bibinfo{author}{\bibfnamefont{G.}~\bibnamefont{Giudici}},
  \bibinfo{author}{\bibfnamefont{M.}~\bibnamefont{Dalmonte}}, \bibnamefont{and}
  \bibinfo{author}{\bibfnamefont{M.~A.} \bibnamefont{Rajabpour}},
  \bibinfo{journal}{Phys. Rev. B} \textbf{\bibinfo{volume}{100}},
  \bibinfo{pages}{155122} (\bibinfo{year}{2019}),
  \urlprefix\url{https://link.aps.org/doi/10.1103/PhysRevB.100.155122}.

\bibitem[{\citenamefont{Mendes-Santos et~al.}(2020)\citenamefont{Mendes-Santos,
  Giudici, Fazio, and Dalmonte}}]{Mendes_Santos_2020}
\bibinfo{author}{\bibfnamefont{T.}~\bibnamefont{Mendes-Santos}},
  \bibinfo{author}{\bibfnamefont{G.}~\bibnamefont{Giudici}},
  \bibinfo{author}{\bibfnamefont{R.}~\bibnamefont{Fazio}}, \bibnamefont{and}
  \bibinfo{author}{\bibfnamefont{M.}~\bibnamefont{Dalmonte}},
  \bibinfo{journal}{New Journal of Physics} \textbf{\bibinfo{volume}{22}},
  \bibinfo{pages}{013044} (\bibinfo{year}{2020}),
  \urlprefix\url{https://doi.org/10.1088%2F1367-2630%2Fab6875}.

\bibitem[{\citenamefont{Zhang et~al.}(2020)\citenamefont{Zhang, Calabrese,
  Dalmonte, and Rajabpour}}]{zhang2020lattice}
\bibinfo{author}{\bibfnamefont{J.}~\bibnamefont{Zhang}},
  \bibinfo{author}{\bibfnamefont{P.}~\bibnamefont{Calabrese}},
  \bibinfo{author}{\bibfnamefont{M.}~\bibnamefont{Dalmonte}}, \bibnamefont{and}
  \bibinfo{author}{\bibfnamefont{M.~A.} \bibnamefont{Rajabpour}},
  \emph{\bibinfo{title}{Lattice bisognano-wichmann modular hamiltonian in
  critical quantum spin chains}} (\bibinfo{year}{2020}), \eprint{2003.00315}.

\bibitem[{\citenamefont{Peschel}(2003)}]{Peschel_2003}
\bibinfo{author}{\bibfnamefont{I.}~\bibnamefont{Peschel}},
  \bibinfo{journal}{Journal of Physics A: Mathematical and General}
  \textbf{\bibinfo{volume}{36}}, \bibinfo{pages}{L205} (\bibinfo{year}{2003}),
  \urlprefix\url{https://doi.org/10.1088%2F0305-4470%2F36%2F14%2F101}.

\bibitem[{\citenamefont{Peschel and Eisler}(2009)}]{Peschel_2009}
\bibinfo{author}{\bibfnamefont{I.}~\bibnamefont{Peschel}} \bibnamefont{and}
  \bibinfo{author}{\bibfnamefont{V.}~\bibnamefont{Eisler}},
  \bibinfo{journal}{Journal of Physics A: Mathematical and Theoretical}
  \textbf{\bibinfo{volume}{42}}, \bibinfo{pages}{504003}
  (\bibinfo{year}{2009}),
  \urlprefix\url{https://doi.org/10.1088%2F1751-8113%2F42%2F50%2F504003}.

\end{thebibliography}

\end{document}